\documentclass[fleqn,usenatbib]{mnras}

\usepackage[T1]{fontenc}

\DeclareRobustCommand{\VAN}[3]{#2}
\let\VANthebibliography\thebibliography
\def\thebibliography{\DeclareRobustCommand{\VAN}[3]{##3}\VANthebibliography}

\usepackage{graphicx}	
\usepackage{amsmath}	
\usepackage{amssymb}	

\usepackage{newtxtext,newtxmath}

\title[Evidence for a dynamic corona in MAXI~J1820+070]{Evidence for a dynamic corona in the short-term time lags of black hole X-ray binary MAXI~J1820+070}

\author[Bollemeijer et al.]{Niek Bollemeijer,$^{1}$\thanks{E-mail: n.a.bollemeijer@uva.nl}
Phil Uttley,$^{1}$
Arkadip Basak,$^{1}$
Adam Ingram,$^{2,3}$
Jakob van den Eijnden,$^{4}$
    \newauthor{Kevin Alabarta,$^{5}$
    Diego Altamirano,$^{6}$
    Zaven Arzoumanian,$^{7}$
    Douglas J.K. Buisson,$^{8}$
    Andrew C. Fabian,$^{9}$}
        \newauthor{Elizabeth Ferrara,$^{7,10}$
        Keith Gendreau,$^{11,12}$
        Jeroen Homan,$^{13}$
        Erin Kara,$^{14}$
        Craig Markwardt,$^{7}$}
            \newauthor{Ronald A. Remillard,$^{14}$
            Andrea Sanna,$^{15}$
            James F. Steiner,$^{16}$
            Francesco Tombesi,$^{17,18,19,10,12}$
            Jingyi Wang,$^{14}$}
                \newauthor{Yanan Wang$^{20,6}$
                and Abderahmen Zoghbi$^{10,21,11}$}
\\
\\
$^{1}$Anton Pannekoek Institute for Astronomy, Amsterdam, Science Park 904, NL-1098 NH, The Netherlands\\
$^{2}$Department of Physics, Astrophysics, University of Oxford, Denys Wilkinson Building, Keble Road, Oxford OX1 3RH, UK\\ 
$^{3}$School of Mathematics, Statistics and Physics, Newcastle University, Herschel Building, Newcastle upon Tyne, NE1 7RU, UK\\
$^{4}$Department of Physics, University of Warwick, Coventry CV4 7AL, UK\\
$^{5}$Center for Astro, Particle and Planetary Physics, New York University Abu Dhabi, PO Box 129188, Abu Dhabi, United Arab Emirates\\
$^{6}$School of Physics and Astronomy, University of Southampton, Southampton, SO17~1BJ, UK\\
$^{7}$X-Ray Astrophysics Laboratory, NASA Goddard Space Flight Center, Code 662, Greenbelt, MD 20771, USA\\
$^{8}$Independent\\
$^{9}$Institute of Astronomy, University of Cambridge, Madingley Road, Cambridge CB3 0HA, UK\\
$^{10}$Department of Astronomy, University of Maryland, College Park, MD, 20742, USA\\
$^{11}$CRESST II, NASA Goddard Space Flight Center, Greenbelt, MD 20771\\
$^{12}$NASA Goddard Space Flight Center, Greenbelt, MD 20771, USA\\
$^{13}$Eureka Scientific, Inc., 2452 Delmer Street, Oakland, CA 94602, USA\\
$^{14}$MIT Kavli Institute for Astrophysics and Space Research, MIT, 77 Massachusetts Avenue, Cambridge, MA 02139, USA\\
$^{15}$Dipartimento di Fisica, Universita` degli Studi di Cagliari, SP Monserrato-Sestu km 0.7, I-09042 Monserrato, Italy\\
$^{16}$Center for Astrophysics, Harvard \& Smithsonian, 60 Garden St, Cambridge, MA 02138, USA\\
$^{17}$Physics Department, Tor Vergata University of Rome, Via della Ricerca Scientifica 1, 00133 Rome, Italy
\\
$^{18}$INAF – Astronomical Observatory of Rome, Via Frascati 33, 00040 Monte Porzio Catone, Italy\\
$^{19}$INFN - Rome Tor Vergata, Via della Ricerca Scientifica 1, 00133 Rome, Italy\\
$^{20}$National Astronomical Observatories, Chinese Academy of Sciences, 20A Datun Road, Beijing 100101, China\\
$^{21}$HEASARC, Code 6601, NASA/GSFC, Greenbelt, MD 20771\\
}

\date{Accepted XXX. Received YYY; in original form ZZZ}

\pubyear{2023}

\begin{document}
\label{firstpage}
\pagerange{\pageref{firstpage}--\pageref{lastpage}}
\maketitle

\begin{abstract}
In X-ray observations of hard state black hole X-ray binaries, rapid variations in accretion disc and coronal power-law emission are correlated and show Fourier-frequency-dependent time lags. On short ($\sim$0.1~s) time-scales, these lags are thought to be due to reverberation and therefore may depend strongly on the geometry of the corona. Low-frequency quasi-periodic oscillations (QPOs) are variations in X-ray flux that have been suggested to arise because of geometric changes in the corona, possibly due to General Relativistic Lense-Thirring precession. Therefore one might expect the short-term time lags to vary on the QPO time-scale. We performed novel spectral-timing analyses on NICER observations of the black hole X-ray binary MAXI~J1820+070 during the hard state of its outburst in 2018 to investigate how the short-term time lags between a disc-dominated and a coronal power-law-dominated energy band vary on different time-scales. Our method can distinguish between variability due to the QPO and broadband noise, and we find a linear correlation between the power-law flux and lag amplitude that is strongest at the QPO frequency. We also introduce a new method to resolve the QPO signal and determine the QPO-phase-dependence of the flux and lag variations, finding that both are very similar. Our results are consistent with a geometric origin of QPOs, but also provide evidence for a dynamic corona with a geometry varying in a similar way over a broad range of time-scales, not just the QPO time-scale. 
\end{abstract}




\section{Introduction}
\label{sec:intro}
Black hole X-ray binaries (BHXRBs) are stellar mass black holes that accrete matter from an orbiting star via an accretion disc. The resultant spectral shape of the disc emission in the X-rays is that of a multi-temperature blackbody \citep{Shakura_Sunyaev_1973,Novikov_1973}. Some fraction of the radiated power takes the form of X-ray power-law emission, thought to be produced by Compton upscattering of disc photons by an optically thin cloud of hot electrons close to the black hole, known as the corona \citep{Sunyaev_1979}. The spectrum of this component is well described by a power-law with a high-energy ($\sim$100~keV) cut-off (e.g. \citealt{Motta_2009}), but the overall shape and geometry of the corona is as yet unknown. Different models have been proposed, with some assuming the corona to be a layer above the surface of the disc close to the black hole \citep{Galeev_1979,Svensson_1994}. The corona could also be a hot flow region between the (thin) disc and the black hole or the base of an astrophysical jet, or some combination of the above (e.g. \citealt{Markoff_2005, Gilfanov_2010, Poutanen_2018, Marcel_2018, Zdziarski_2021}). It should be noted that recent X-ray polarisation results favour a horizontal geometry in the plane of the accretion disc \citep{Krawczynski_2022}. The third major component of the X-ray spectrum of BHXRBs originates from the reflection of high energy photons on the disc. The reflection spectrum consists mainly of a continuum `Compton hump' peaking around $\sim$30~keV and emission lines, of which the Fe~K$\alpha$ line is usually the strongest \citep{George_1991,Matt_1997,Bambi_2021}.

Despite high data quality and advanced modelling, the spectral constraints on the properties of the corona are often degenerate. Adding time-domain information can break these degeneracies and provide better constraints on e.g. the geometry of the corona \citep{Uttley_2014}.
There are several timing properties that can be measured in light curves (see e.g. \citealt{Uttley_2014} for a detailed review of X-ray spectral-timing techniques). In this paper, we will focus on the time lag between variations in two X-ray energy bands (0.5--1~keV and 3--10~keV) for the BHXRB MAXI~J1820+070 in its hard state. By calculating the phase of the mean Fourier cross-spectrum of two light curves in different energy bands, it is possible to measure the time lag between correlated variations in both light curves on different time-scales \citep{Nowak_1999}. Lags between different energy bands are found in many XRBs. By probing the strong broadband-noise variability observed in the hard state in a soft and a hard band, dominated respectively by emission from the disc and from the corona, two main types of lags are distinguished \citep{Uttley_2011,DeMarco_2015,Kara_2019}. At low Fourier frequencies, hard photon variability lags behind variations in the softer X-rays (typically by tenths of a second), so these are called hard lags. At higher Fourier frequencies, variations in the soft band are often found to arrive later than those in the hard band (typically by 1-10~ms), a signature known as soft lags.

Plots of lag vs energy demonstrate that most of the low-frequency hard lag can be associated with a delay between variations of the (lagging) coronal power-law and (leading) disc blackbody emission \citep{Uttley_2011}. This delay can be naturally explained if the variability is produced by mass accretion fluctuations propagating inwards through the disc \citep{Lyubarskii_1997} to a central coronal region. 

When comparing two harder energy bands, a smaller energy-dependent hard lag is seen within the power-law emission, which has been variously attributed to Comptonisation light-travel delays in the corona \citep{Kazanas_1997,Bellavita_2022} or jet \citep{Reig_2003,Kylafis_2008,Wang_Yanan_2020}, accretion fluctuations propagating through a coronal hot flow \citep{Kotov_2001,Arevalo_2006,Veledina_2013} or coronal heating vs. seed photon delays in response to accretion fluctuations propagating through the disc to the corona \citep{uttley2023large}. When measuring lags between 3--10~keV, which consists mainly of coronal power-law emission, and 0.5--1~keV, which is dominated by disc emission but also contains coronal emission, it is important to realise that the observed lags arise due to a combination of several mechanisms. The relative strength of different spectral components can influence the measured lag values.

To explain the soft lags between disc and coronal emission observed at higher Fourier frequencies, different mechanisms have been proposed. \citet{Mushtukov_2018} argue that outward propagation of mass accretion fluctuations can cause soft lags and \citet{Kawamura_2023} show that inwardly propagating fluctuations can lead to soft lags if the inner region has a higher low energy break in its spectrum than the outer region, which could happen in the hard-intermediate state (HIMS). \citet{Uttley_2011}, \citet{DeMarco_2015} and \citet{Alston_2020} interpret soft lags as evidence of reverberation. In their view, photons are emitted by the corona in all directions, and some will be thermally reprocessed at lower energies by the disc. Due to e.g. the light travel time and the reprocessing timescale between both components, variations in the soft band are observed later than those in the hard band, to produce soft lags. Reverberation lags should be highly dependent on coronal geometry, which influences what fraction of hard photons is reprocessed by the disc and which also determines the size of the lags and at what frequency the lags switch from hard to soft.\footnote{The lags between two power-law dominated energy bands remain positive or go to zero at high frequencies and do not show a turnover to soft lags.} \citet{Kara_2019} found that time lags on short time-scales ($>3$~Hz) change during the 2018 outburst of black hole X-ray binary MAXI~J1820+070, which is the source that is also being analysed in this paper. They interpreted these changes in lags as being related to changes in the vertical extent of the corona. 

If high-frequency time lags depend on coronal geometry, they can potentially be used to study a different phenomenon observed in many black hole X-ray binaries, called quasi-periodic oscillations or QPOs (for recent reviews of QPOs, see \citealt{Motta_2016} and \citealt{ingram2020review}). Several subtypes of QPO are distinguished. During the hard state, low-frequency QPOs of type C, which have a centroid frequency ranging from a few mHz up to $\sim$10~Hz, are typically present as a narrow peak in the power spectrum of the light curve. \citet{Wang_2022} showed that the high-frequency lag behaviour is tightly linked to the type-C QPO frequency in MAXI~J1820+070 and other BHXRBs.

\begin{figure}
    \centering
    \includegraphics[width = \columnwidth]{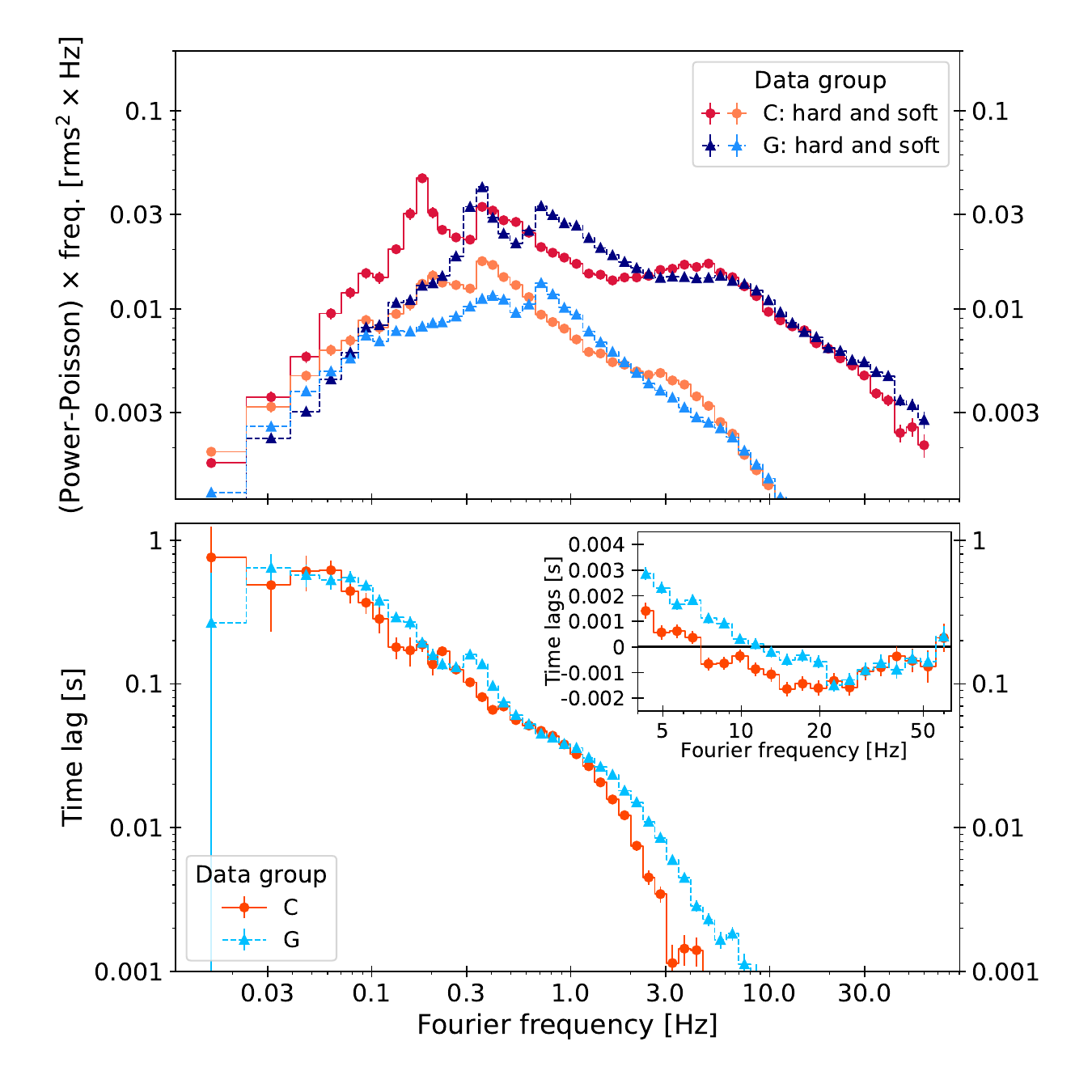}
    \caption{
    Examination of the mean power spectrum and time lags for all observations with a similar QPO frequency combined. We group ObsIDs into 10 data sets (A through J) with increasing QPO frequency (see Table \ref{ObsID_datasets} for more details.) Here we demonstrate the lags and PSD for two data groups, C and G, where the mean QPO frequency is 0.18 Hz and 0.34 Hz, respectively. The QPO is stronger in the hard band and a harmonic signal at twice the QPO fundamental frequency is clearly visible in both data sets. The lower panel shows the time lags for the same data sets. The short-term time lags around 4-20~Hz change during the outburst. The inset plot shows the soft lags on a linear scale.
    }
    \label{ps_hs_lags_3sets}
\end{figure}

The changes in both the power spectra and the short-term time lags of MAXI J1820+070 are visualised in Fig. \ref{ps_hs_lags_3sets}, which shows power spectra and lag-frequency spectra for two different data groups. The upper panel of Fig. \ref{ps_hs_lags_3sets} shows power spectra for two subsets of data with different QPO-frequencies, with prominent peaks in the hard band power spectra demonstrating the presence of QPOs. In the power spectrum, QPO signals often consist of two clear peaks: one at the fundamental frequency and one at around twice that frequency, which is called the second harmonic. We will refer to the latter as the harmonic from now on. Variations at the fundamental and harmonic frequencies are linked in phase, and together they create a specific QPO waveform, which can be measured (see e.g. \citealt{Ingram_2015,de_Ruiter_2019,Nathan_2022}).

A number of competing models have been proposed to explain low-frequency QPOs. Various models suppose that the QPO corresponds to intrinsic luminosity changes, linked to characteristic frequencies or instabilities in the accretion flow or jet, e.g. corrugation modes \citep{Tsang_2013}, the accretion-ejection instability \citep{Varniere_2012}, or an instability in the jet \citep{Ferreira_2022}. However, a number of observational results support the idea that the QPO is primarily due to a geometric variation in the appearance of the corona. Evidence that the QPO corresponds to a varying coronal geometry is provided by the binary system inclination dependence of QPO phase lags \citep{van_den_Eijnden_2016_inc} and amplitudes \citep{Motta_2015,Heil_2015}, as well as QPO phase-resolved spectroscopy \citep{Ingram_2016_IronK,Ingram_2017}. The geometric interpretation is supported by the study of high energy QPO lags \citep{Ma_2021} and the variability spectrum of the QPO (see e.g. \citealt{Sobolewska_2006}, \citealt{Axelsson_2013} and \citealt{Gao_2023} with \textit{Insight-HXMT} data for MAXI~J1820+070). The foremost explanation for these geometric variations is the relativistic precession model \citep{Stella_1999,Schnittman_2006,Fragile_2007,Ingram_2009}. In the model, the hot inner flow and associated corona and/or jet precess due to the General Relativistic Lense-Thirring effect \citep{Lense_Thirring_1918}. Recent General Relativistic Magnetohydrodynamical (GRMHD) simulations also demonstrate the effect (see e.g. \citealt{Liska_2018}, \citealt{liska2019phase}, \citealt{Musoke_2022} and \citealt{Bollimpalli_2022}). 

If the coronal geometry changes during a QPO cycle, light travel times and reflection fractions will vary as well, and so we can expect that high-frequency time lags should also vary over the QPO cycle. Connecting QPO phase and short-term time lags is therefore an important way to test the geometric interpretation of QPOs. In this paper, we analyse the variability of high-frequency time-lags in a hard state BHXRB on different time-scales, to determine whether there is evidence for variability in the lags that may be linked to the QPO or other types of variation.

To search for variations in the short-term time lags on the QPO time-scale, high quality data is required in order to ensure that the signal at high Fourier frequencies is not overwhelmed by observational noise. Also, the studied source should show both broadband noise variability and QPOs. All these conditions are met by observations of black hole X-ray binary MAXI~J1820+070 in its hard state with the \textit{Neutron Star Interior Composition ExploreR} (NICER) \citep{Gendreau_2016}, obtained during the 2018 outburst. We describe these data and our basic methods in Section~\ref{data_subsection}.

In Section~\ref{methods_section} we present our analysis and results. We first look at how the lags change with flux, which is highly variable on a range of time-scales. We then use Fourier filtering methods to show how the correlation between lag and flux depends on the variability time-scale, to show that the strongest relation arises at the QPO time-scale. We also introduce a new method to identify the QPO phase at each point in time in a light curve, using a combination of Fourier filtering and interpolation, to measure how the short-term time lags depend on QPO phase. In order to test our new method, we used a more established method, pioneered by \citet{Ingram_2015} and applied to many XRBs by \citet{de_Ruiter_2019}, to see if it returned similar results in terms of QPO waveform reconstruction, which is presented in the Appendix, along with a description of a simulation approach to check the validity of our results.  We end with a discussion of the implications of our results for different models for the QPO and the coronal geometry.

\section{Observations and methods}
\label{data_subsection}
We made use of the extensive observations by NICER of the low mass black hole X-ray binary MAXI~J1820+070 (known in the optical as ASASSN-18ey, \citealt{Torres_2019}). NICER has excellent timing capabilities and can observe bright sources with minimal instrumental deadtime and no spectral distortion such as due to pileup. It is located on the International Space Station (ISS). MAXI~J1820+070 was one of the brightest X-ray sources ever observed during its outburst in 2018, which led to an extraordinarily rich data set (see e.g. \citealt{Buisson_2019,Kara_2019,Fabian_2020,Homan_2020,Axelsson_2021,Wang_2021,Zdziarski_2021,De_Marco_2021}; \citealt{You_2021,You_2023}).

\begin{table*}
    \centering
    \begin{tabular}{c|c|c|c|c}
        \hline
        Data set & ObsIDs & Selected exposure time~[s] & QPO frequency~[Hz] & Flux range 3-10 keV [cts/s]\\
        \hline\hline

        A & 126, 127, 129, 130 & 6,080 & 0.07-0.12 & 852 - 1932 \\
        B & 131, 132, 133, 134 & 10,560 & 0.12-0.16 & 839 - 1861\\
        C & 135, 136, 137, 138, 139 & 14,080 & 0.16-0.20 & 856 - 1835 \\
        D & 181, 182, 183, 184, 185 & 5,248 & 0.20-0.24 & 243 - 524\\
        E & 140, 141, 142, 176, 177, 178, 180, 186 & 20,416 & 0.24-0.28 & 509 - 1047 \\
        F & 143, 144, 169, 171, 173, 174, 175, 179, 187, & 21,888 & 0.28-0.32 & 545 - 1108\\
        G & 145, 146, 158, 159, 160, 161 & 32,384 & 0.32-0.36 & 620 - 1236 \\
          & 162, 167, 168, 170, 172, 188  && \\
        H & 147, 165, 166 & 9,152 & 0.36-0.40 & 675 - 1320\\
        I & 148, 151, 157, 163, 164, 189 & 26,560 & 0.40-0.44 & 560 - 1092\\
        J & 149, 150, 152, 153, 154, 155, 156 & 12,224 & 0.44-0.51 & 758 - 1452\\
        \hline

    \end{tabular}
    \caption{The combinations of ObsIDs used in different parts of this paper. Also, the total selected exposure time and the QPO frequency range for each data set of ObsIDs is shown. More information on the individual ObsIDs can be found in Table \ref{ObsID_list}. The rightmost column shows the range of hard fluxes (lowest and highest flux bin count rates) obtained by creating 5 flux bins of 0.25 s slices within each 64 s segment, as explained in Subsection~\ref{subsection_flux_lags}.}
    \label{ObsID_datasets}
\end{table*}

We analysed data collected between 2018 April 11 (ObsID 1200120126, MJD 58219) and 2018 June 28 (ObsID 1200120189, MJD 58297), while MAXI~J1820+070 was in the hard state. During this part of the outburst, the QPO fundamental frequency varied between 0.077 and 0.51~Hz. For the remainder of this paper, data sets will be referred to by the last three digits of their ObsIDs. The full ObsID number is constructed by placing 1200120 before these three digits. One ObsID refers to observations that are obtained during one Earth day. The list of ObsIDs and their total observation time can be found in Table \ref{ObsID_list}. Because data set D, corresponding to QPO frequencies between 0.2 and 0.24~Hz, only contains five relatively short observations with lower count rates and weak QPOs, results for this frequency range typically have large error bars and should be interpreted with caution.

All ObsIDs were reprocessed using the \texttt{nicerl2} pipeline from NASA's HEASoft (v6.28) \citep{FTOOLS_2014}, version 6, released on 2020 July 22. Good Time Intervals (GTIs) were applied to all ObsIDs used. Any GTI shorter than 64~s was excluded. For GTIs longer than 64~s, only an integer number of 64~s segments was used and partially filled segments were discarded. In ObsIDs 130 and 131, from 2018 April 16 and 17, there were several GTIs in which an unusual flux decrease was observed. This was caused by the robot arm on ISS crossing in front of NICER during observations\footnote{\url{https://heasarc.gsfc.nasa.gov/docs/nicer/timelines/nicer_significant_events.html}}, and could be identified through its characteristic shadow pattern crossing the 56 FPMs. Affected time intervals were omitted.

We created light curves with a time resolution of 1/128~s, resulting in a Nyquist frequency of~64~Hz. Light curves were made for two energy bands. The soft band is defined to be 0.5--1~keV, where disc emission dominates for disc temperatures between $\sim0.2$ and $0.3$~keV, as observed in the hard state of MAXI~J1820+070 \citep{Dzielak_2021,Wang_2021}. The hard band ranges from 3--10~keV, where the power-law component originating from the corona is the main contributor. To investigate how the short-term time lags vary on a time-scale of seconds, all 64~s segments are subdivided into 0.25~s slices. The choice for a certain slice length over which the lags are calculated is based on two properties. First, the available Fourier frequency range is restricted to high frequencies for short slices. The lowest Fourier frequency measured for these short slices is close to the Poisson noise dominated region, resulting in low signal-to-noise ratios (SNR). However, later in our analysis, we want to be able to distinguish well between different QPO phases. Since the QPO frequency ranges from $\sim$0.08 to 0.5~Hz for the data analysed here, corresponding to periods between 2 and 12~s, the slice length should be a fraction of these periods. Balancing between these two arguments, we use 0.25~s slices, so the lowest frequency measured is 4~Hz. At this frequency, Poisson noise does not dominate in the energy bands used. The frequency range over which the time lags are determined is 4 to 20~Hz. The measured lags in this frequency range arise due to a combination of hard lags at lower frequencies and soft lags at higher frequencies. Both the hard and the soft lags show strong long-term evolution \citep{Kara_2019,De_Marco_2021,Wang_2022}. Higher frequencies do not add much extra signal, because Poisson noise dominates the fast variability. To keep our results model-independent, we do not investigate the hard and soft lags in the 4 to 20~Hz range separately. Instead, we show that the average 4--20~Hz time lags probe the evolution of the soft lags well (see e.g. Figs. \ref{tlag_freq_alldata} and \ref{tlag_flux_alldata}).

The Fourier cross-spectrum of the hard and the soft band is calculated for all simultaneous hard and soft band light curve slices. In order to obtain good signal-to-noise, it is necessary to average the cross-spectra of many slices. The way we select which slices to average together is discussed in the Section~\ref{methods_section}. To obtain the phase lag between both bands for those slices, the argument of the average of the cross-spectrum over the 4--20~Hz range for all included light curve slices is then calculated. The phase lag can be converted to a time lag by dividing by $2\pi\bar{\nu}$, where $\bar{\nu}$ is the mean frequency of the range for which the lag was calculated. In our analysis, the mean frequency is $\bar{\nu}=12$~Hz.

We also calculate the errors on the phase lags following \citet{Uttley_2014}, using 
\begin{equation}
    \Delta\phi(\nu_j) = \sqrt{\frac{1-\gamma^2(\nu_j)}{2KM\gamma^2(\nu_j)}},
\end{equation}
where $\gamma^2(\nu_j)$ is the raw coherence between both light curves for frequency bin $\nu_j$, $K$ is the number of Fourier frequencies over which the lags are averaged (e.g. for 4--20~Hz and 0.25~s light curve slices, $K=5$) and $M$ is the number of slices used to form the average cross-spectrum. The phase lag errors are converted to time lag errors by dividing by $2\pi\bar{\nu}$. For a higher SNR, synonymous with smaller error bars on the time lags, we combined multiple observations to obtain the cross-spectrum, and applied the described methods to all available data. The QPO frequency and with it the shape of the power spectrum evolves considerably during the outburst. The averaged cross-spectrum over a frequency range is weighted by the power in each frequency bin, so a changing power-spectral shape can influence the lag measurement, even if the lags themselves do not change. We also perform the same analysis on ten separate subsets of data, consisting of ObsIDs with similar QPO frequencies, which are listed in Table \ref{ObsID_datasets}.

\section{Data analysis and results}
\label{methods_section}

\begin{figure*}
    \includegraphics[width=149mm]{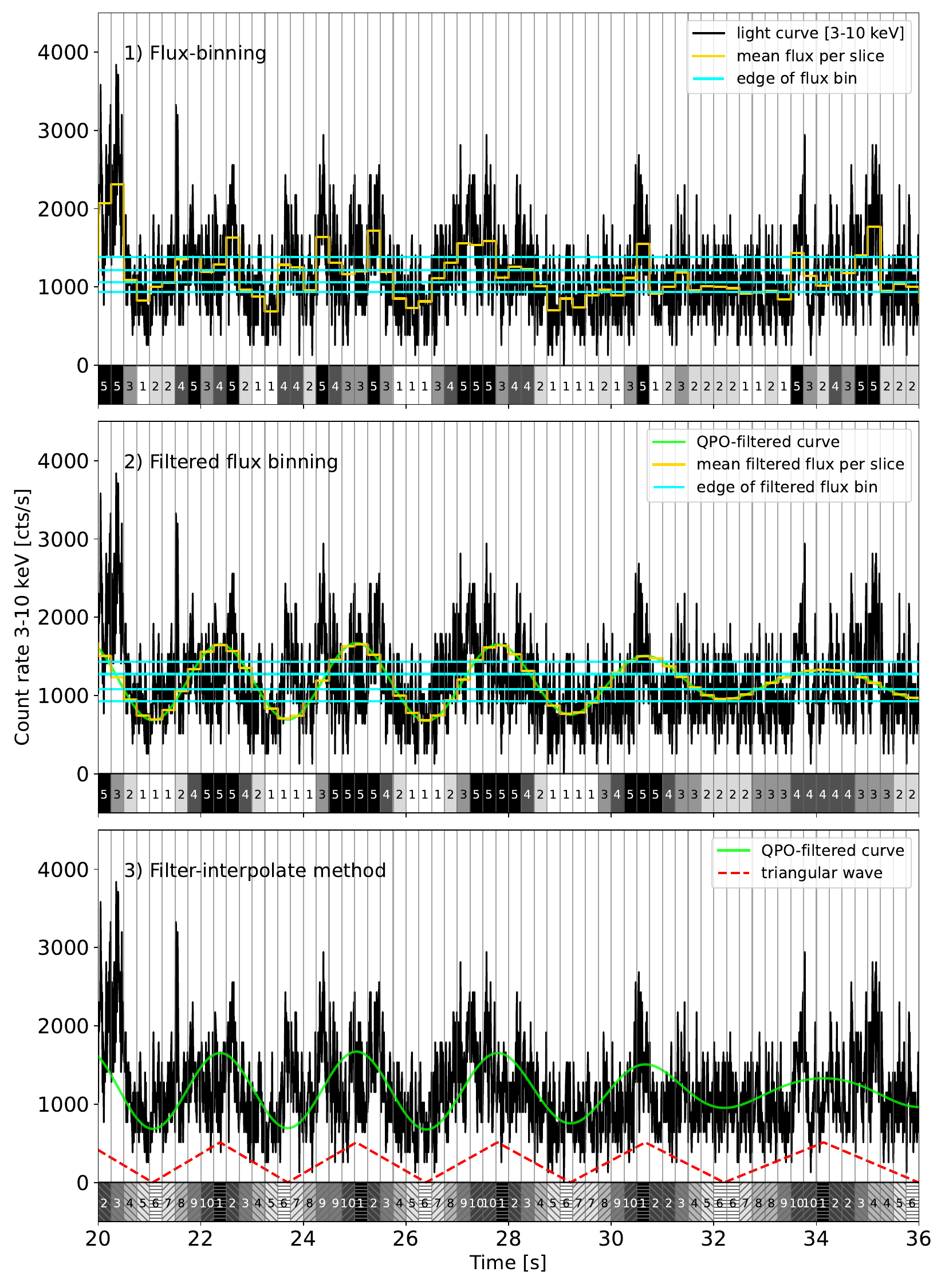}
    \caption{To explain the three methods we use to analyse the behaviour of the short-term time lags, the three panels show how the 0.25 s slices are combined in different ways for each method with data from ObsID 145. The 3-10 keV light curve is shown in black and the vertical lines show the edges of each 0.25 s slice. The numbers and shades below the light curves each correspond to a single flux bin in the upper panels, while the lower panel also contains diagonal hatching to show the different phase bins. The cross-spectra of simultaneous 3-10 keV and 0.5-1 keV light curve slices with the same shade and number are averaged to calculate the lags of each flux or phase bin. The edges of the flux bins are calculated for each individual 64 s segment and are visible as horizontal cyan lines in panel 1) for unfiltered flux binning and in panel 2) for QPO-filtered flux binning. For the green QPO-filtered curve, it is clear that both the phase and amplitude shift over time. Because the figure only shows 16~s of a 64~s segment, not all flux or phase bins are filled to the same extent, but there is an equal number of slices in each bin for the entire segment. In panel 2 and 3, the amplitude of the filtered light curves and the edges of the filtered flux bins were multiplied by a factor of 5 for clarity, so the actual flux difference between individual filtered flux bins is much smaller than shown in the figure.}
    \label{expl_3methods}
\end{figure*}
Our goal is to investigate how the short-term time lags change on a time-scale of seconds and how they are connected to QPOs, which might shed light on changing coronal-geometry on short time-scales. We can measure the short-term lags by averaging the cross-spectra obtained from many 0.25 s light curve slices, but in order to determine their relation to specific variability time-scales or the QPO, we consider three different methods for `binning' the slices together for cross-spectrum averaging purposes. The three methods are schematically explained in Fig. \ref{expl_3methods}, which has one panel for each method, showing 16~s of hard band light curve and indicating how 0.25~s slices are combined. We briefly introduce each method below, before providing further details and results in separate sub-sections:
\begin{enumerate}
\item First, we investigate whether the lags depend on the flux variability on a time-scale between 0.25 and 64~s, which are the lengths of a slice and of a segment. Both QPO and broadband noise variability play an important role on these time-scales. Short 0.25~s slices are grouped together based on their relative flux within 64~s segments, so any variability on longer time-scales (e.g. as the source evolves during the outburst) does not affect the flux-binning. The upper panel of Fig. \ref{expl_3methods} shows how the different slices are combined, which is discussed more extensively in Subsection~\ref{subsection_flux_lags}. 
\item In the framework of a geometric origin of QPOs, we expect the lag to vary more on the QPO time-scale than on other time-scales. We therefore devised a method using Fourier filtering to isolate the variability on a specific time-scale, using only the flux variations on that time-scale to bin slices according to their relative flux. We can then probe how variability on the QPO time-scale influences the short-term time lags and compare it to the effects of variability on other time-scales. The filtered flux binning is shown schematically in the middle panel of Fig.~\ref{expl_3methods} and explained in more detail in Subsection~\ref{subsection_lag_var_timescales}. 
\item Finally, we present a method to resolve the QPO signal and follow how the short-term time lags evolve with QPO phase. The main goal is to investigate whether changes in flux and lag happen simultaneously or that there is a delay between both, which could indicate different types of geometric change. We call it the filter-interpolate (FI) method, as shown in the lower panel of Fig. \ref{expl_3methods} and presented in Subsection~\ref{subsection_qpophase_sawtoothmethod}. 
\end{enumerate}

In all of the methods described in this section, the lags are obtained from the average cross-spectrum of a large number of short slices. The short-term time lags evolve during the outburst and thus each group or bin will consist of slices with different intrinsic lags. The absolute value of the average lags therefore does not have much physical meaning.
However, we can still compare the lag values in different bins to measure a relation between lag and flux, as all the methods only take into account variability within a 64~s segment.

To test our methods and to be able to distinguish systematic effects from results that are intrinsic to the data, we simulated `null hypothesis' light curves with constant lags based on \citet{TK95}, to which we added a method to simulate QPOs. The methods we used to simulate data are discussed in more detail in Appendix \ref{sim_appendix}.

\subsection{Short-term lag response to flux variations on time-scales up to 64~s}
\label{subsection_flux_lags}
We combine different 0.25~s slices according to their mean flux by calculating the mean flux of each slice and sorting the obtained values for each 64~s segment. Within each 64~s segment, the 51 slices with the lowest mean fluxes are assigned to the same flux bin, the next 51 slices to the next bin and so on, until we have 5 different flux bins. This is illustrated in the upper panel of Fig.~\ref{expl_3methods}. We decided to create 5 flux bins to have sufficient signal-to-noise in each bin, also for subsets of observations, while retaining a median flux bin. Because each segment consists of 256 slices of 0.25~s and we can only assign 255 of them to a flux bin, we exclude the last 0.25~s of each (unsorted) segment. The binning process is performed for all 64~s segments and the respective flux bins of each segment are combined to create 5 flux bins, each consisting of thousands of light curve slices. It is important to note that the mean flux value that is used to combine different slices can be either the hard or the soft flux, i.e. 3--10~keV or 0.5--1~keV, as explained in the previous sections. We carried out the analysis and show the results for both cases, but Fig.~\ref{expl_3methods} only shows an example for binning on 3-10 keV flux. Because dividing the 0.25 s slices over the five flux bins is only done according to the relative flux within each 64 s segment, the long term evolution of the flux does not influence the binning.

We calculate the short-term time lags between the hard and soft energy band for each flux bin as described in Section~\ref{data_subsection}. Once the mean short-term time lags for the five distinct flux bins are obtained, models can be fitted to the five data points to quantify the relation between the lags and the flux. We fit a constant and a linear model to determine whether there is a significant change of short-term time lags with flux.

\begin{figure}
    \centering
    \includegraphics[width=\columnwidth]{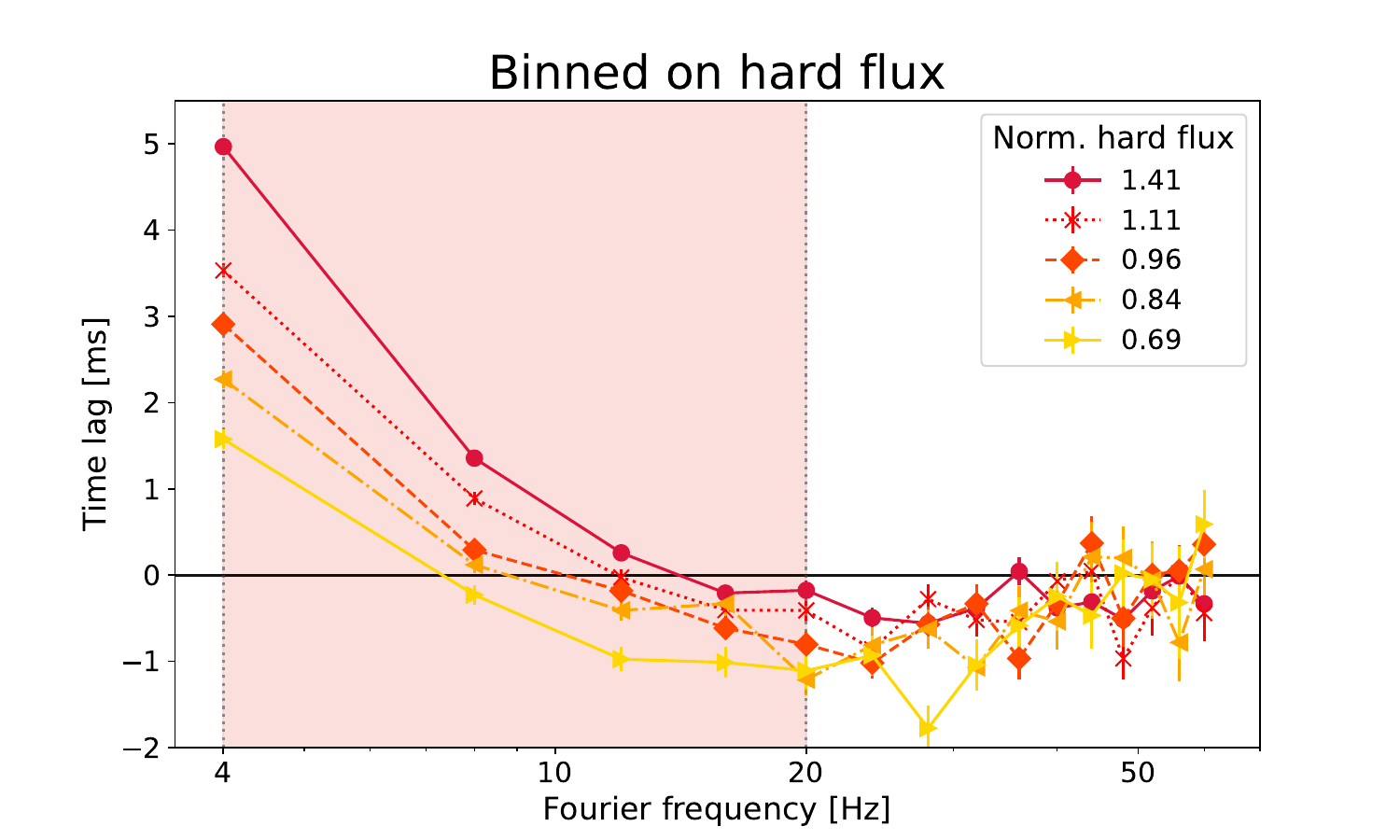}
    \includegraphics[width=\columnwidth]{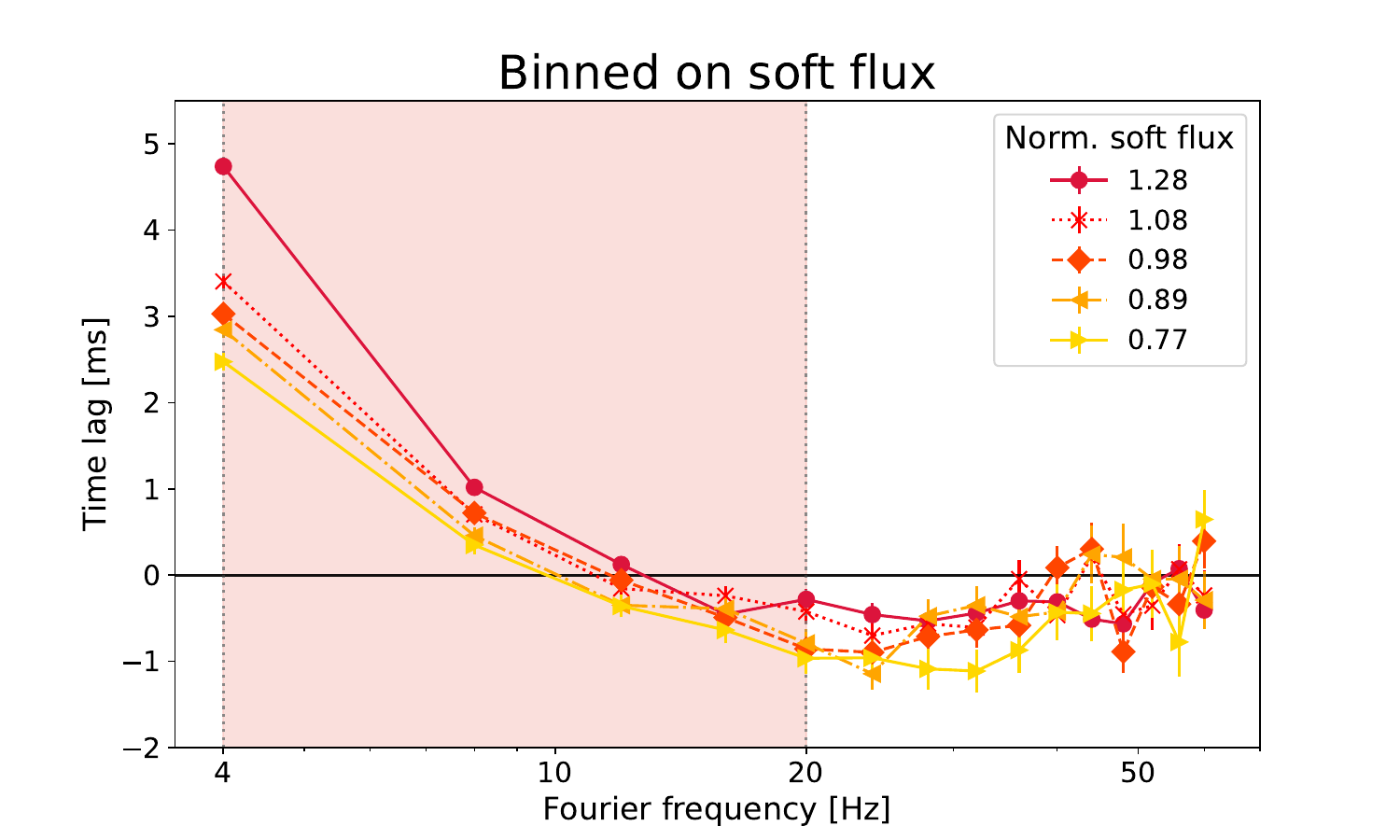}
    \caption{The upper panel shows the lag-frequency spectra of the five different hard flux bins, while the lower panel shows the same for soft flux bins. It is clear that the amplitude of the soft lag decreases and the frequency at which the lags switch from hard to soft increases with flux and that the effect of binning on hard flux is larger than for the soft flux. The 4--20~Hz range over which the lags are calculated for Fig. \ref{tlag_flux_alldata} is highlighted.}
    \label{tlag_freq_alldata}
\end{figure}

\begin{figure}
    \centering
    \includegraphics[width=\columnwidth]{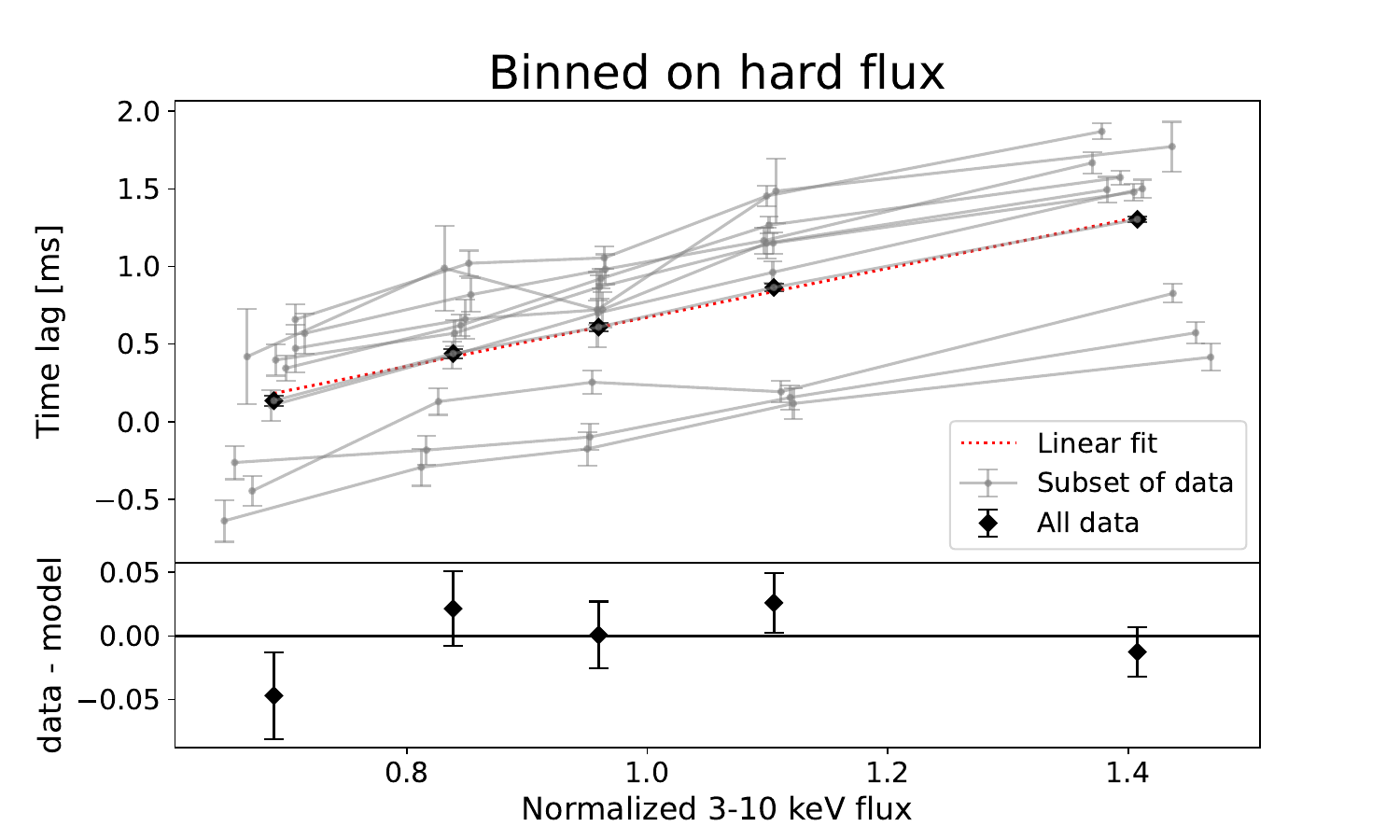}
    \includegraphics[width=\columnwidth]{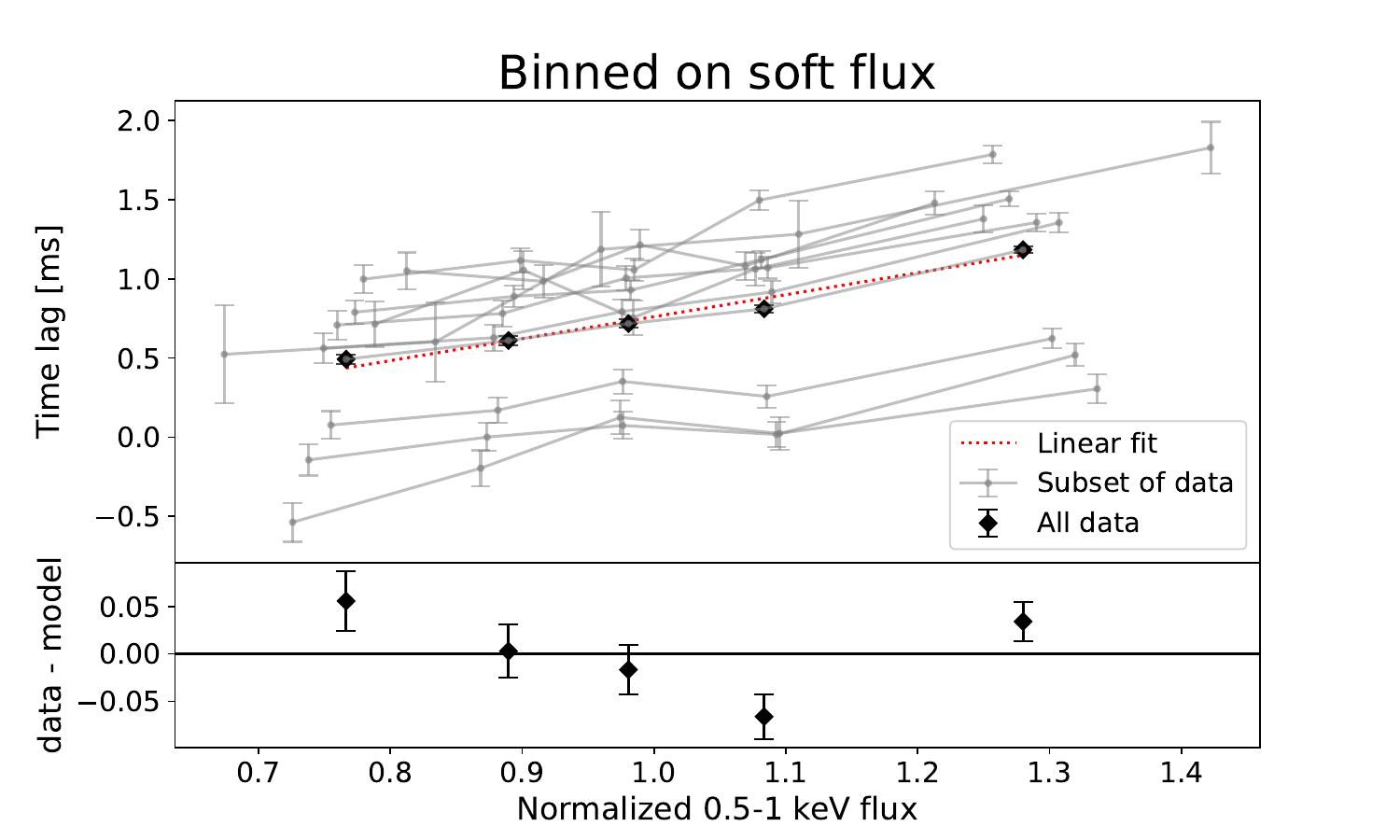}
    \caption{The upper panel shows the normalised hard flux versus the mean 4 to 20~Hz time lags between the hard and the soft band for all included ObsIDs as black diamonds. The lags and flux were calculated for 0.25~s slices. The lighter curves show the result of flux binning for the different subsets of data, grouped by their QPO frequency and shown in Table \ref{ObsID_datasets}. All subsets show the same trend with hard flux, while the mean time lag differs between observations. The lower panel shows the same results for binning on the soft 0.5--1~keV flux.}
    \label{tlag_flux_alldata}
\end{figure}

In Fig. \ref{tlag_freq_alldata}, the lag-frequency spectra for the five different hard and soft flux-binned 0.25~s light curve slices are shown for all data. The lag-frequency dependence clearly evolves, as the cross-over frequency from hard to soft lags increases and the maximum soft lag decreases with flux. The net hard lag in the 4-20 Hz range therefore increases with flux, as is illustrated in Fig. \ref{tlag_flux_alldata}, where we show the relation between the mean (normalised) hard and soft flux and the short-term (4--20~Hz) time lags. The normalisation involves dividing the flux of each slice by the average flux of all slices in its segment. A linear fit and its corresponding residuals are also shown. It is clear that the linear fit describes the data reasonably well for the hard flux binned data, which is demonstrated by a goodness of fit (p-value) of 0.26 with Pearson's $\chi^2$ test. The slope of the linear fit for binning on hard flux is $1.58\pm 0.05$~ms. 

The lower panels of Figs. \ref{tlag_freq_alldata} and \ref{tlag_flux_alldata} show the behaviour of the short-term time lags when binning on soft flux. Although the overall trend is the same, several discrepancies between the soft and hard flux-binned data can be discerned. The goodness of fit of the linear function to the soft flux binned data is 0.003, indicating a more complex relation. Also, the ranges of both normalised flux and time lags are smaller when binning on soft flux. The slope of the linear fit is slightly shallower when binning on soft flux instead of hard flux with a value of $1.39\pm0.14$~ms. From the combination of these properties, we conclude that the short-term time lags are connected more closely to the hard band variations than to the soft band variations. Further results are only shown for binning on the hard band, even though similar effects can be distinguished when binned on soft flux. Since variability in both hard and soft light curves is correlated, it is not surprising that binning on either of them returns qualitatively similar effects.

The linear lag vs. flux relation is not only seen when combining all observations, but also in the different subsets of data as presented in Table \ref{ObsID_datasets}. The results for the subsets of data are shown as the grey lines in Fig. \ref{tlag_flux_alldata}. The time lag range covered by binning on flux is very similar for all subsets of data. The subsets with higher QPO frequencies, corresponding to the lines with a more positive average lag in Fig. \ref{tlag_flux_alldata}, display a reduced range in flux, which leads to slightly steeper slopes. 


\subsection{Time-scale dependence of short-term lag response to flux variations}
\label{subsection_lag_var_timescales}
In Subsection~\ref{subsection_flux_lags}, we show that the short-term time lags vary linearly with flux on time-scales of 0.25--64~s, but it is unclear whether the relation is driven by the QPO or broadband noise variability, or both. To distinguish between QPO and broadband noise variability, we use a method that can successfully extract the QPO signal from a light curve. We follow e.g. \citet{van_den_Eijnden_2016} and \citet{Press_1992} and apply an optimal filter to the Fourier transform (FT) of each light curve segment. To obtain the optimal filter, we first fit five Lorentzians to the mean power spectrum of all light curve segments in a single ObsID. Three broad Lorentzians account for the broadband noise and two narrow Lorentzians are centered at the QPO fundamental and harmonic frequencies. Poisson noise is accounted for by adding a constant to the fit. The optimal filter consists of the narrow QPO fundamental Lorentzian function divided by the sum of all five Lorentzian functions, which leaves us with a filter that closely resembles the shape of the QPO peak in the power spectrum (see Fig. 2 in \citet{van_den_Eijnden_2016} for an example of an optimal filter). All frequencies outside twice the FWHM of the narrow fundamental Lorentzian are excluded. Because the QPO strength varies during the outburst and the fits return rather different QPO widths, we assume a Q-factor of 8 for all ObsIDs. $Q=\nu_\textsc{qpo}/\textsc{fwhm}$ and it is a measure of how many cycles on average the QPO signal stays in phase with itself. $Q\sim8$ or higher for typical type-C QPOs observed at $>3$~keV \citep{ingram2020review}.

The inverse FT of the filtered FT returns a smooth and filtered light curve containing variations on the QPO time-scale. This is illustrated in the middle panel of Fig. \ref{expl_3methods}, where the QPO filtered light curve and the mean value of that filtered light curve in each slice are shown in green and yellow, respectively. Slices of the original black light curve are combined according to the value of the filtered light curve, as illustrated by the shades and numbers below the x-axis. Slices with the same number belong to the same filtered flux bin. Only variability on the QPO time-scale is taken into account when creating the five different bins. The mean cross-spectrum between simultaneous slices in the (unfiltered) hard and soft band and the corresponding short-term time lags in the 4 to 20~Hz range are calculated for each of the five filtered flux bins.

To compare the results of filtering on QPO and non-QPO frequencies, we can move the centroid frequency of the Lorentzian fitting the QPO to other frequencies, while keeping the broad Lorentzians the same. We then only take into account variability from a chosen, (non-)QPO time-scale to combine 0.25 s slices from the original (unfiltered) hard and soft light curves. The rest of the filtering process is identical to the QPO case. We always used an optimal filter with $Q = 8$ at both QPO and non-QPO frequencies. An alternative, applying a top-hat filter with the same width, returned similar results. It is important to note that even if we use an optimal filter at the QPO frequency, the broadband noise will still contribute to the filtered light curve, as the optimal filter cannot distinguish between overlapping QPO and broadband noise variability.

\begin{figure}
    \centering
    \includegraphics[width=\columnwidth]{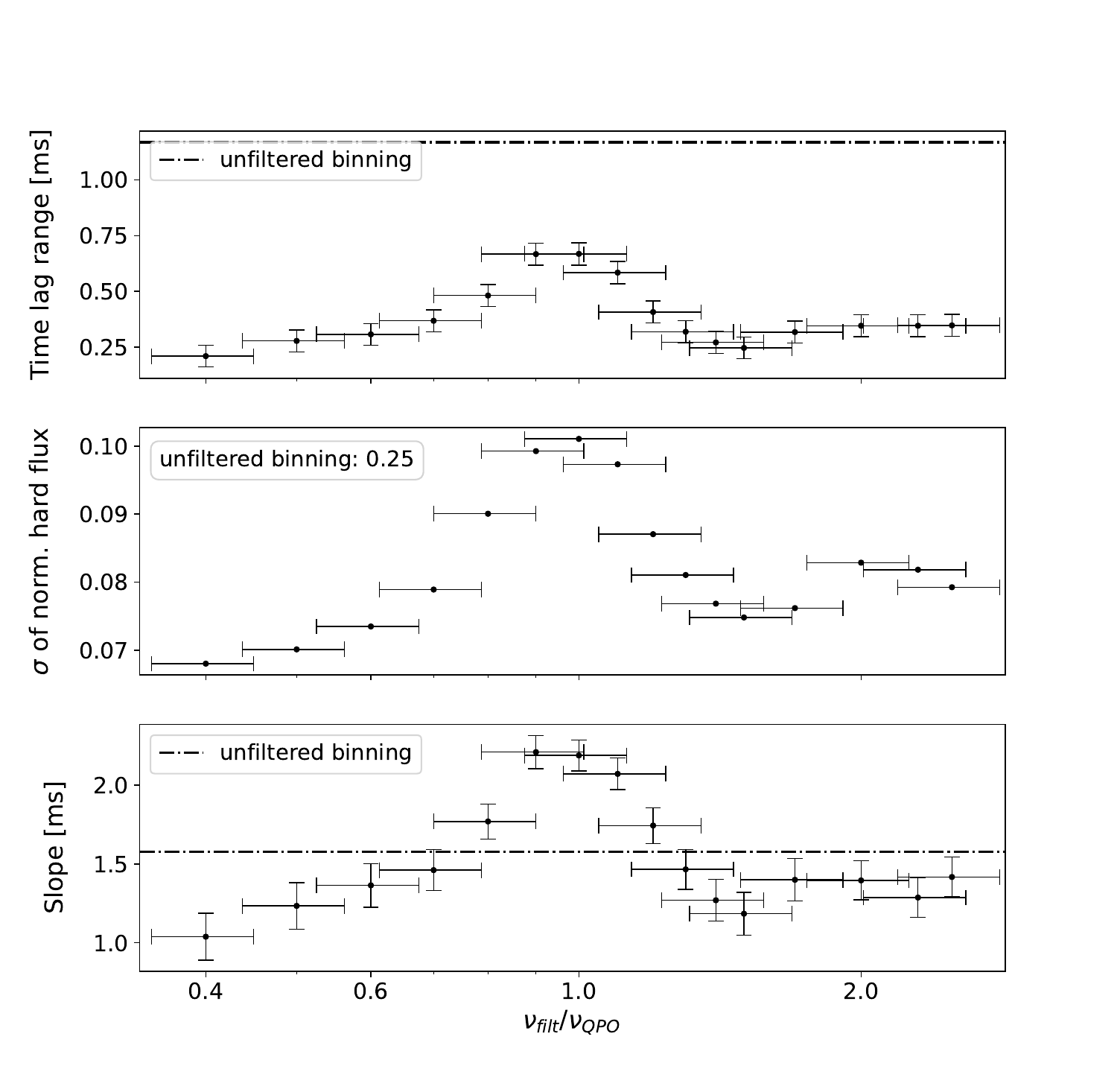}
    \caption{The result of measuring and fitting the time lags versus flux after binning 0.25~s light curve slices on filtered flux for different Fourier frequency filters for all data combined. On the x-axis, the filtering frequency divided by the QPO frequency of each included ObsID is shown and the horizontal error bars indicate the width of the optimal filter that was used. The upper panel shows the time lag range, defined as the difference in time lags between the lowest and highest flux bin. The middle panel shows the standard deviation of the normalised filtered flux bin values, which indicates the variability on the filtered time-scale. The standard deviation $\sigma=0.25$ for unfiltered flux binning. This is much larger than for filtered flux binning, as filtering removes a large part of the variability. In the lower panel, the slope of a linear fit to the lag-flux relation (see Fig. \ref{tlag_flux_alldata}) is shown. The QPO clearly stands out in all three panels.}
    \label{3plot_lag_flux}
\end{figure}

By shifting the centroid frequency of the optimal filter, choosing to include or exclude the QPO frequency, we can separate out the QPO time-scales from the broadband noise continuum and map the dependence of the short-term time lags on flux variations at different time-scales. However, since the flux variability amplitude is time-scale-dependent, the lags are likely to be influenced by the flux range covered by filtering on a given range of Fourier frequencies. To determine the strength of the connection between flux and lags on different time-scales, we fit a linear function to the short-term time lag as a function of hard flux, similar to what is visible in Fig. \ref{tlag_flux_alldata}. We use the slope of the lag-flux relation to determine whether the QPO has a different relation to variations in short-term time lags than the broadband noise. The result of the analysis for all data combined is shown in Fig. \ref{3plot_lag_flux}. 

The upper panel shows the difference in the time lags between the lowest and highest (filtered) flux bin. There is a clear peak at the QPO frequency. In principle, the presence of the peak could be explained by the fact that the flux variability is large at the QPO frequency, as is visible in the middle panel. If there is a simple linear relation between flux and time lag, large flux variability leads to an increase in time lag range. However, in that case, the slope of the relation is not expected to change. In the lower panel of Fig. \ref{3plot_lag_flux}, it is clear that the slope of the relation is steeper when using a light curve filtered on the QPO frequency than when using filters on non-QPO frequencies or the unfiltered flux as the binning property. This suggests that there is a profound connection between the QPO and the variations of the short-term time lags.

Fig. \ref{3plot_lag_flux} shows 16 filter frequencies between 0.4 and 2.5 times the QPO frequency. Filter frequencies above this range lead to strong biases in the lag measurements, which are also observed in simulated light curves with a constant lag, while these biases are much weaker for lower filter frequencies. An explanation of the bias can be found in Appendix \ref{app_rednoise}.

\subsection{QPO phase-resolved time lags: the filter-interpolate method}
\label{subsection_qpophase_sawtoothmethod}

To further understand how short-term time lags and the QPO are related, we designed the filter-interpolate (FI) method for estimating the QPO phase at a given time. The first steps are the same as those used for measuring short-term time lag variations on different time-scales in Subsection~\ref{subsection_lag_var_timescales}. After creating the QPO filtered light curve by applying the same optimal filter to the hard band, we determine the time values of the extrema of the filtered curve. With these points in time, a triangular wave is created with value 1 at the maxima and value -1 at the minima. A linear interpolation connects these points, hence the name of the method. Interpolation is necessary because both the QPO frequency (or equivalently, phase) {\it and} amplitude vary from one cycle to the next \citep{van_den_Eijnden_2016}. The FI method is shown visually in the lower panel of Fig. \ref{expl_3methods}. The original filtered line is shown as the solid green curve, while the result of the interpolation is visible as the dashed, red triangular wave in the same figure. The vertical lines indicate the start and stop times of each light curve slice. Each slice is assigned a QPO phase based on the value of the triangular wave in the center of the slice and the sign of its slope, which means we assume that the QPO frequency does not change during half a QPO cycle. Each QPO phase corresponds to a distinct shade and number below the x-axis. The inclusion of the sign of the slope of the interpolation is the main difference between the FI method and the method described in Subsection~\ref{subsection_lag_var_timescales} and shown in panel 2 of Fig. \ref{expl_3methods}. With the sign of the slope we can distinguish between rising and falling parts of the QPO signal and resolve the QPO phase. Another difference between both methods is that the FI method only takes into account the phase of the filtered curve, while binning on filtered flux depends on the amplitude of the filtered curve in each slice.

\begin{table*}
    \centering
    \begin{tabular}{c|c|c|c|c|c}
    \hline
    \multicolumn{1}{c}{\textbf{Filter:}} &
    \multicolumn{2}{c}{\textbf{QPO frequency}} & 
    \multicolumn{1}{c}{\vline}& 
    \multicolumn{2}{c}{\textbf{1.5 $\times$ QPO frequency}} \\
    \hline
    & \textbf{Norm. flux}   &  \textbf{Time lags} & \vline & \textbf{Norm. flux} & \textbf{Time lags} \\
    \hline 
    Offset & 1 (fixed) & $0.773 \pm 0.013$~ms&\vline& 1 (fixed) &$0.84 \pm 0.012$~ms\\
    Fundamental amplitude ($A_{\rm f}$)  &$0.1372 \pm 0.0006$&$0.319\pm 0.016$~ms&\vline&$0.0977 \pm 0.0005$& $0.121 \pm 0.016$~ms\\
    Amplitude ratio ($R_{\rm amp}$) &$0.186 \pm 0.005$&$0.22 \pm 0.06$&\vline&$0.074 \pm 0.007$ &$0.40 \pm 0.17$ \\
    $\phi_{\rm{f}}$ [rad] &$0.315 \pm 0.004$&$0.12 \pm 0.05$&\vline&$0.34 \pm 0.005$& $0.16 \pm 0.12$ \\
    $\Psi/\pi$ &$0.320 \pm 0.005$&$0.33 \pm 0.05$&\vline&$0.206 \pm 0.014$& $0.3 \pm 0.3$ \\
    \hline
    \end{tabular}
    \caption{The fitting parameters for a model consisting of a fundamental and harmonic sine wave to the flux and lag waveforms (equation \ref{equation_qpo_main}) when filtering on the QPO frequency or 1.5 times the QPO frequency, which are shown in Figs. \ref{phlag_waveform1} and \ref{phlag_waveform_1.6QPOf}. The $1\sigma$ errors on the parameters were determined by bootstrapping the 64~s segments of all data 100 times and performing the analysis on the bootstrapped samples. Because the non-QPO lag waveform does not contain a significant harmonic, the amplitude ratio and $\Psi$ are not well-constrained and have large errors.}
    \label{waveform_parameters}
\end{table*}

\begin{figure}
    \centering
    \includegraphics[width=\columnwidth]{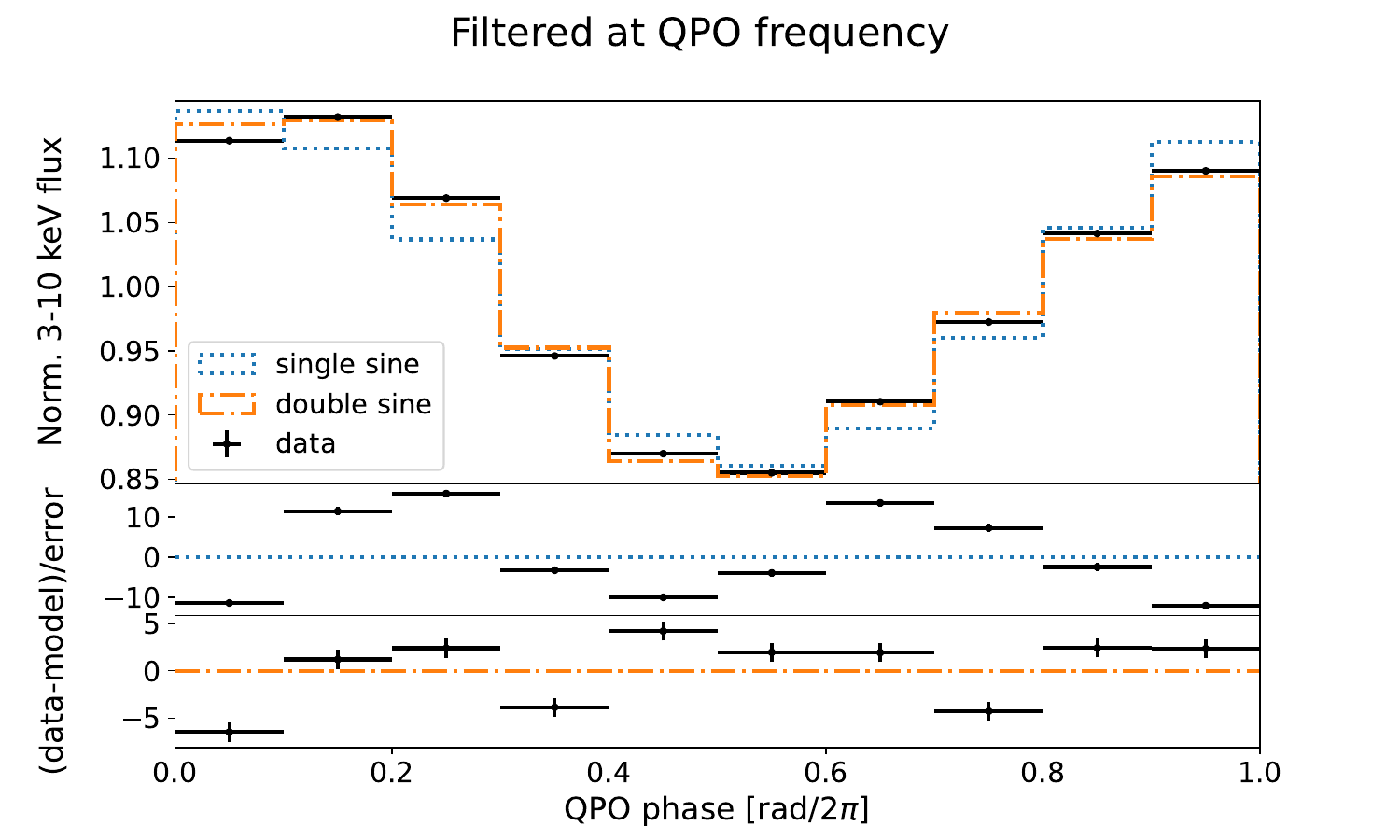}
    \includegraphics[width=\columnwidth]{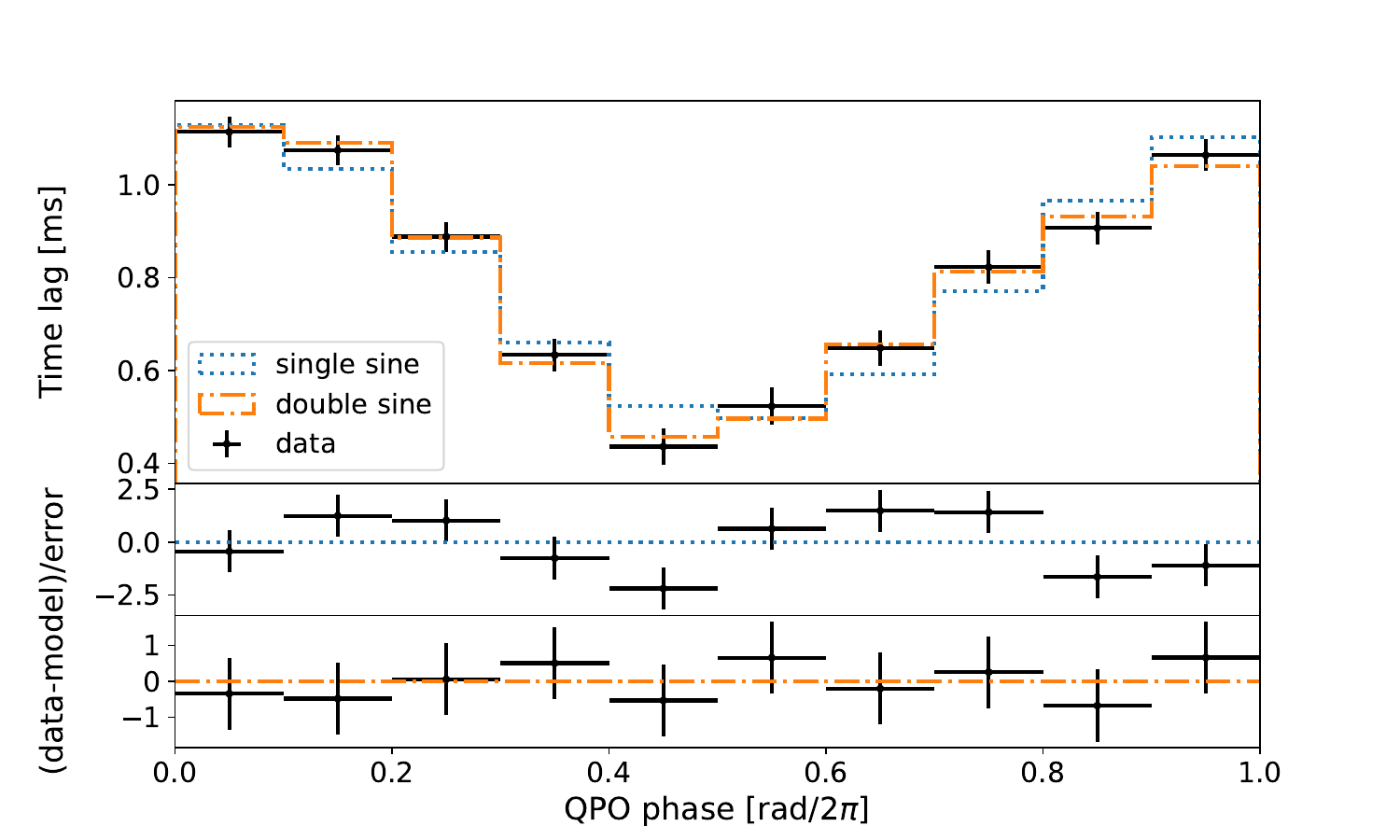}
    \caption{The hard flux and 4--20~Hz time lag (between the hard and the soft band) waveforms for all data, obtained by the FI method at the QPO frequency. Also, a single sine wave and a double sine wave fit are shown. It is clear that the time lag follows the shape of the flux. For both waveforms, the addition of a harmonic signal significantly improves the fit. The error bars on the flux are underestimated.}
    \label{phlag_waveform1}
\end{figure}

Next, we can associate a QPO phase with the {\it unfiltered} hard and soft light curve slices so that we can calculate a cross-spectrum for different QPO phase bins. Any light curve slice shorter than $\sim$1/4 of the QPO period can be assigned a QPO phase by the FI method. Longer slices will average out variations due to the QPO. The 0.25~s slices used in this work are not influenced by this effect, as even the highest QPO frequency in the data set ($\sim$0.5~Hz) has a period of about 8 slices. Each hard and soft 0.25~s light curve slice is assigned to one of ten equal-width QPO phase bins and the mean cross-spectrum between the two energy bands for all slices in a phase bin is calculated and used to calculate the short-term (4--20~Hz) time lags. The ten time lag values obtained constitute the time lag waveform, which informs us how the short-term time lags evolve with QPO phase. The flux and lag waveforms for the QPO for all data combined are visible in Fig. \ref{phlag_waveform1}.

When we have the ten QPO phase bins for both the hard flux and the short-term time lags, we can fit a function $f(\Theta)$ (where $\Theta$ is the phase in radians, defined at the central filter frequency) to the waveforms, which is the sum of two cosine waves corresponding to the fundamental and harmonic components:
\begin{equation}
\label{equation_qpo_main}
    f(\Theta) = 1 + A_{\rm f} \cos(\Theta - \phi_{\rm f}) + R_{\rm amp}A_{\rm f}\cos(2[\Theta - \phi_{\rm f} - \Psi]).
\end{equation}
Here, $A_{\rm f}$ is the fundamental amplitude, $R_{\rm amp}$ is the ratio of the harmonic and the fundamental amplitudes and $\Psi$ is the phase difference that determines the waveform, defined by \citet{Ingram_2015} as
\begin{equation}
\label{equation_psi}
    \Psi=(\phi_{\rm{h}}/2-\phi_{\rm{f}}) \text{ mod } \pi,
\end{equation}
where $\phi_{\rm{f}}$ and $\phi_{\rm{h}}$ are the phase offsets of the fundamental and harmonic respectively. The division by 2 is due to the fact that the frequency of the harmonic is twice the frequency of the fundamental, so its phase changes twice as much for an equal shift in time. With the fit, we can quantify the waveforms, as is discussed in more detail in Appendices \ref{sim_appendix} and \ref{appendix_waveform}. 

\begin{figure}
    \centering
    \includegraphics[width=\columnwidth]{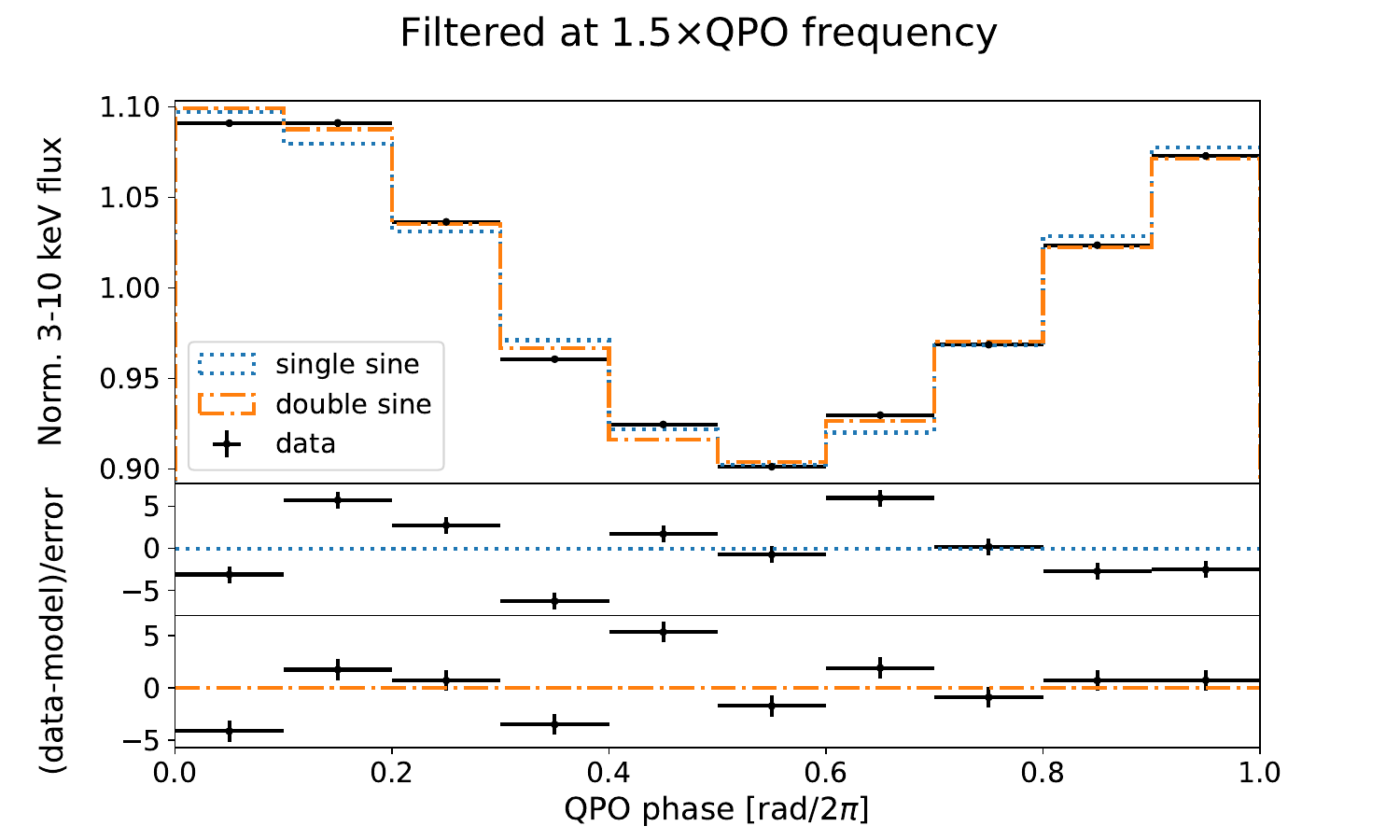}
    \includegraphics[width=\columnwidth]{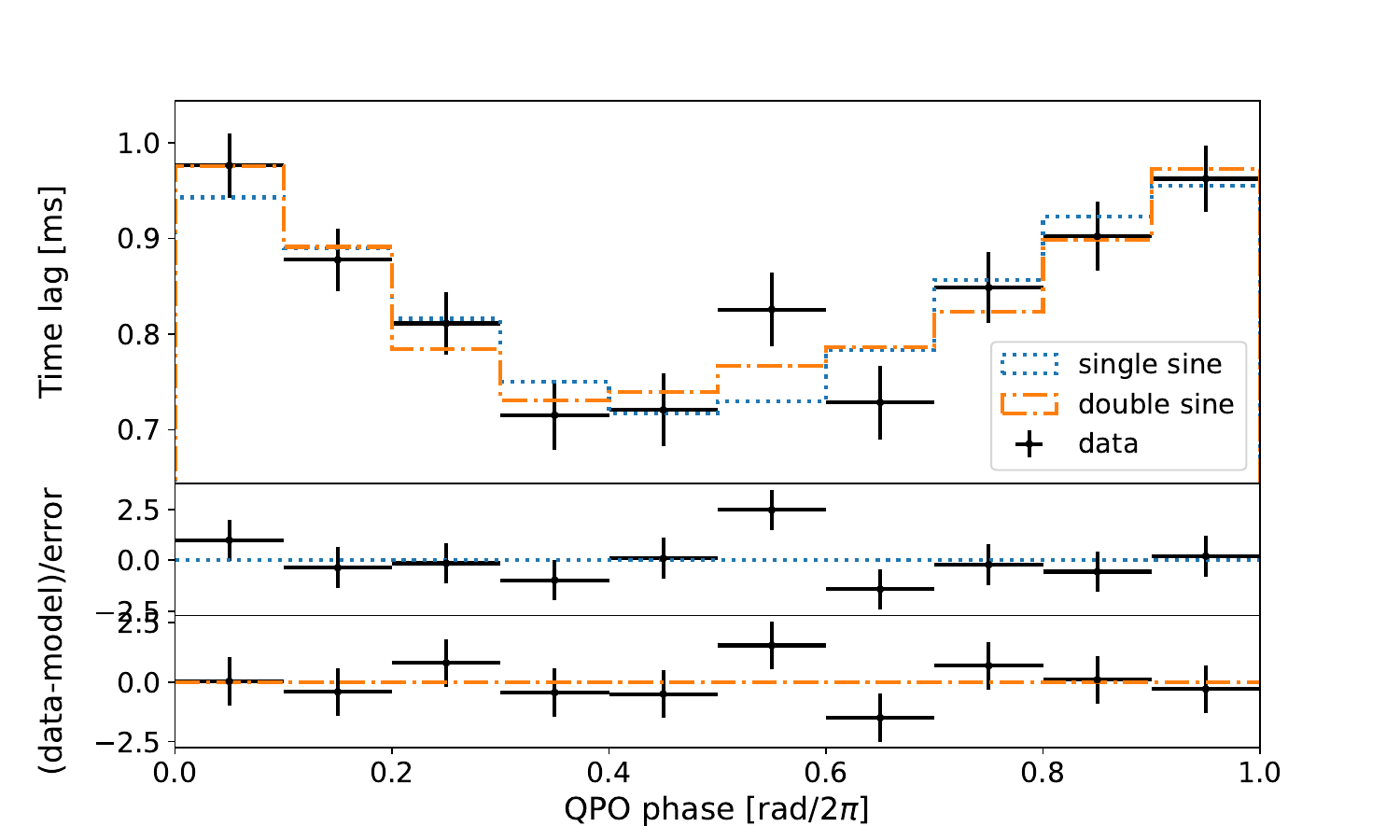}
    \caption{The hard flux and 4--20~Hz time lag (between the hard and the soft band) waveforms for all data, obtained by the FI method at 1.5 times the QPO frequency. The harmonic contribution to the flux waveform is much smaller than when filtering on the QPO frequency (Fig. \ref{phlag_waveform1}) and the harmonic does not improve the fit to the lag waveform significantly.}
    \label{phlag_waveform_1.6QPOf}
\end{figure}

In Fig. \ref{phlag_waveform1}, the hard flux and time lag QPO waveforms are shown for all data combined. We fit both a single (same as equation \ref{equation_qpo_main} with $R_{\rm amp}=0$) and a double sine wave model to the waveforms, integrating the model over each flux bin. It is clear that the flux waveform requires a harmonic signal, at twice the filtering frequency, in order to be fit reasonably well. We used the presence of a harmonic signal in the QPO waveform to test our method, which is described in Appendix \ref{appendix_waveform}. The parameters of the fits are shown in Table \ref{waveform_parameters}. The errors on the flux obtained from counting statistics (simply $1/\sqrt{N_{\rm{counts}}}$, where $N_{\rm{counts}}$ is the number of counts in a flux bin) are too small, as they do not account for correlated errors between phase bins that are introduced by our FI method of phase reconstruction used to bin light curve slices. Therefore we determined the errors on the fitting parameters by bootstrapping 64~s segments of data 100 times and fitting double sine waves to the waveforms obtained from the bootstrapped samples. The standard deviations of the parameter distributions are reported in Table \ref{waveform_parameters}. The errors on the lags are more consistent with the scatter of the data points and we find the lag waveform contains a harmonic signal with a $3.5\sigma$ significance\footnote{The significance is calculated from the difference in $\chi^2$, which decreases from 17.5 for a single sine wave and 7 degrees of freedom (d.o.f.) to 2.2 for a double sine wave with 5 d.o.f.}. The values for the harmonic-fundamental amplitude ratio and phase difference $\Psi$ are consistent with those of the flux waveform, indicating that the short-term lags are tightly connected to the flux.

Analogous to what we show in Subsection~\ref{subsection_lag_var_timescales}, we can investigate both QPO and non-QPO variability and their relation to the short-term time lags by shifting the centroid of the optimal filter. We created optimally filtered hard band light curves for different Fourier frequency ranges and compare the results. The rest of the method is the same as described above, so we can compare the effect of including the QPO in the filtering frequency range.
As an example, Fig. \ref{phlag_waveform_1.6QPOf} shows the hard flux and time lag waveforms for broadband noise at 1.5 times the QPO frequency of each ObsID. As can also be seen in Table \ref{waveform_parameters}, the broadband noise flux waveform contains a harmonic component that is significantly weaker than in the QPO case and probably originates from the broadband noise rms-flux relation, which causes Fourier phases to be correlated between frequencies (e.g. see \citealt{Uttley_2005}). The lag waveform fit does not improve significantly ($1.5\sigma$)\footnote{The $\chi^2$ decreases from 10.5 for 7 d.o.f. to 6.3 for 5 d.o.f.} when including a harmonic signal, which could be due to the relatively large error bars on the lags (due to the lower rms in the flux).

From Table \ref{waveform_parameters} it is clear that the values for the fundamental phase $\phi_{\rm{f}}$ are different for the flux and lag waveforms. The non-zero phase difference\footnote{To avoid confusion: $\Psi$ is the phase difference between the fundamental and harmonic frequencies in a light curve (see Appendix \ref{sim_appendix}), while the phase difference referred to here is between the fundamental of the flux and lag waveform, which we will also call the delay between flux and lag.} indicates that the lags and the flux do not vary simultaneously, but that there is a delay between variations in the flux and in the short-term time lags. To investigate the delay further, we determined the phase difference between the lag and flux waveforms by fitting a single sine wave to both and comparing their phases, so we do not take into account any harmonic signal here. This simplified model is used to avoid degeneracies in the phase caused by the harmonic components, which can compensate for differences in the phase of the fundamental. 

\begin{figure*}
    \centering
    \includegraphics[width =175mm]{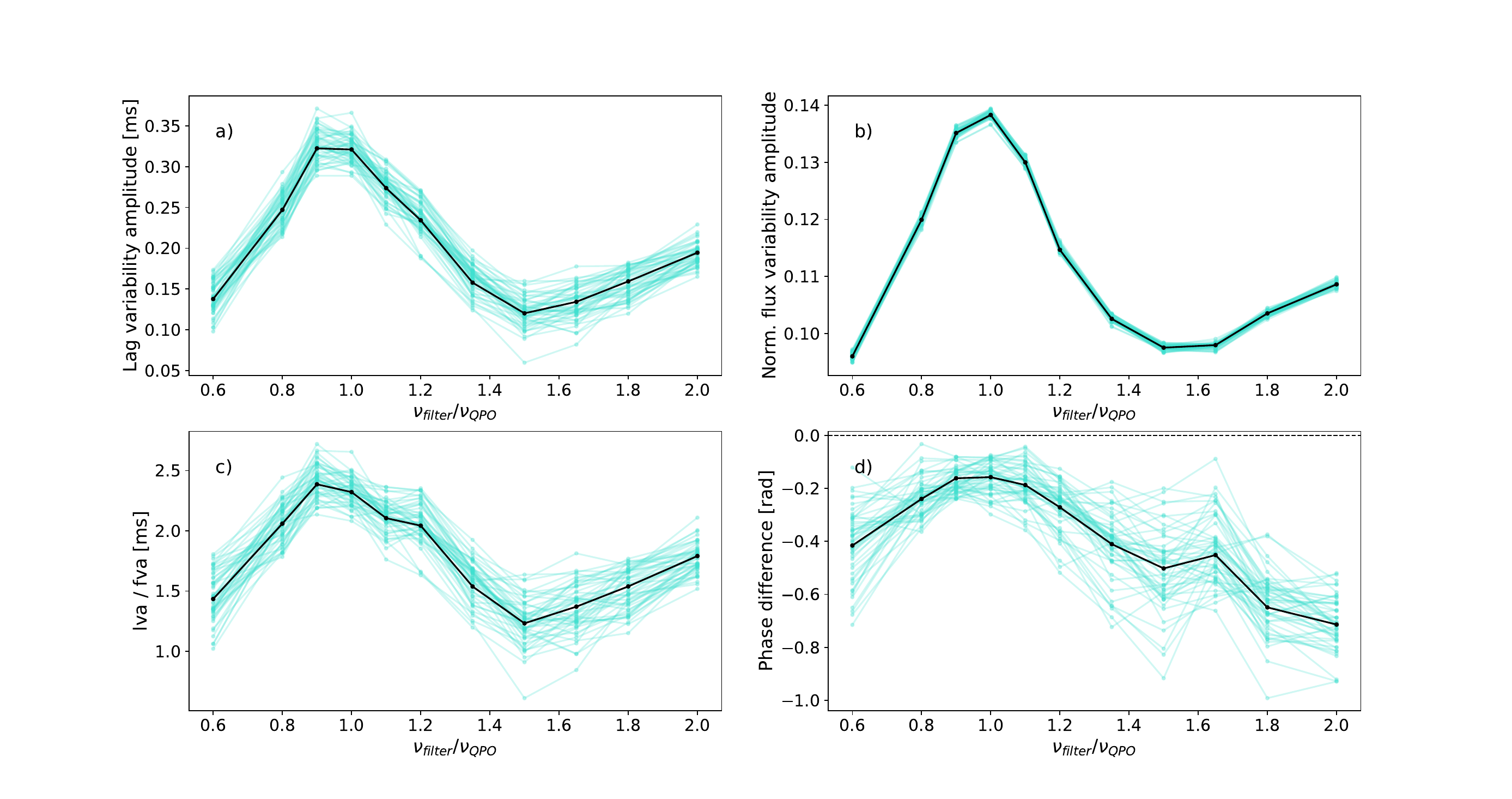}
    \caption{In panel a) the lag variability amplitude (lva) (4--20~Hz between 3--10 and 0.5--1~keV) is shown, in panel b) the flux variability amplitude (fva), the ratio of both is visible in panel c) and panel d) shows the phase difference between the short-term time lag and flux waveforms for all included data for different filter frequencies relative to the QPO frequency of each observation. The turquoise lines show the total of 50 bootstraps that were used to determine the errors, while the solid black lines represent the original data. Bootstrapping was used to account for the errors on different filter frequencies, which are correlated. There is a clear maximum for both variability amplitudes and their ratio when filtering at the QPO frequency. In the lower right panel, a negative phase difference indicates that the flux follows the short-term time lags, which is due to phase leakage also seen in null-hypothesis simulations.}
    \label{filterfreqfig}
    \centering
    \includegraphics[width =175mm]{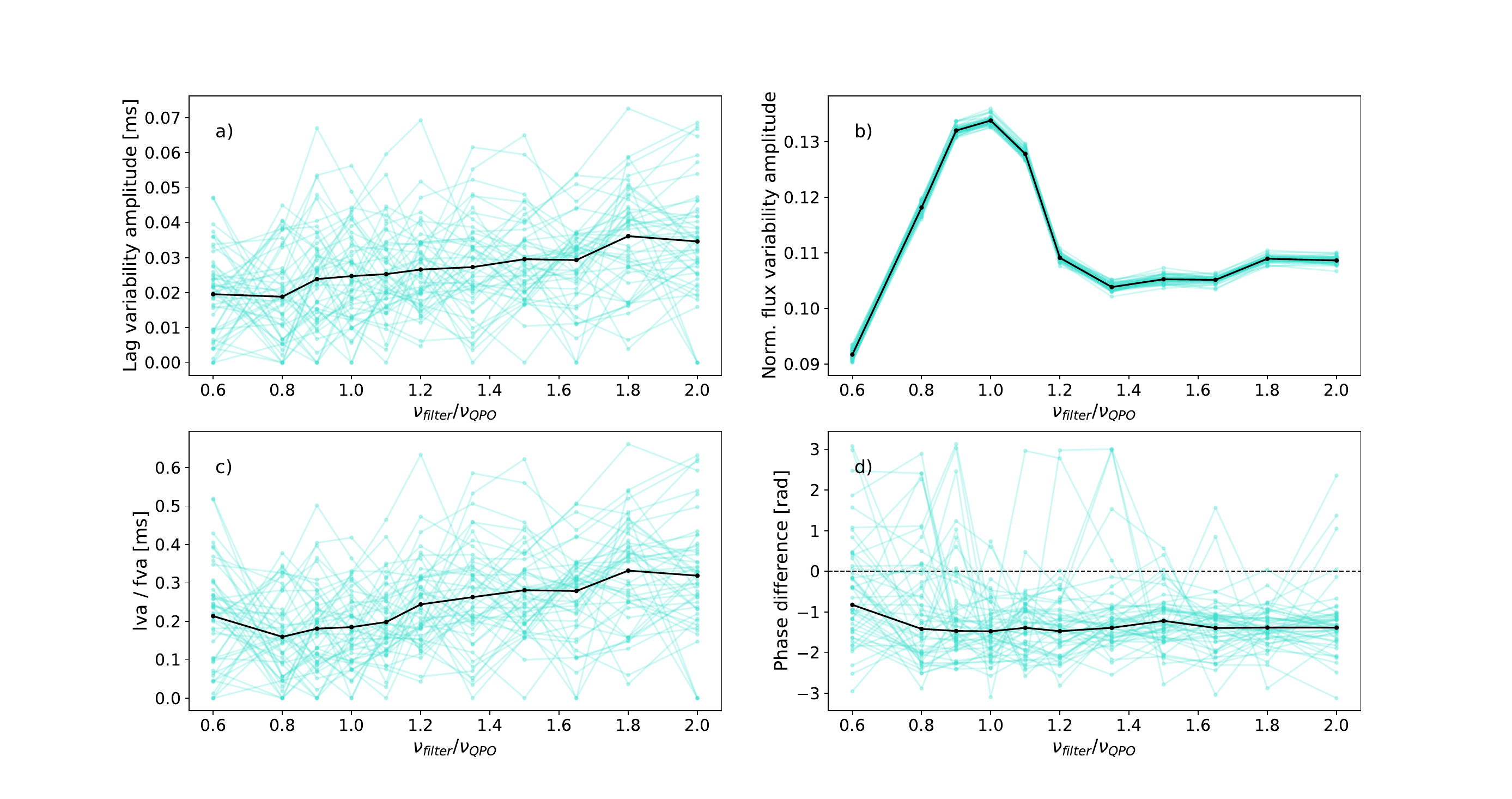}
    \caption{The 4--20~Hz lag variability amplitude (lva) is shown in panel a), panel b) shows the flux variability amplitude (fva), c) their ratio and d) the phase difference between the short-term time lag and flux waveforms for 50 `null-hypothesis' simulations consisting of 1000 segments of 64~s with a QPO frequency of 0.31~Hz. The black lines shows the mean value of the parameters of the 50 simulations. Because there is no intrinsic lag-flux relation in the simulation, the lva is much smaller than in the data. The lag waveform that arises due to Fourier leakage has a slightly larger amplitude for higher filter frequencies and tends to precede the flux waveform by $\sim\pi/2$~rad, as is visible in the lower right panel.}
    \label{filterfreqfig_sim}
\end{figure*}

Fig. \ref{filterfreqfig} shows the lag variability amplitude (lva), flux variability amplitude (fva), their ratio and the phase difference between the lag and the flux waveforms for different filter frequencies. The lva and fva are defined as the amplitudes of the sine functions fitted to the lag and flux waveforms as presented in Figs. \ref{phlag_waveform1} and \ref{phlag_waveform_1.6QPOf}. The errors were determined by bootstrapping the 64~s segments and applying the FI method to the bootstrapped samples. Bootstrapping the 64~s segments accounts for the fact that neighbouring filter frequency results are correlated, which is also visible in the turquoise bootstrapped curves. The lower right panel of Fig. \ref{filterfreqfig} shows that the lag and flux waveforms are almost but not quite simultaneous and there is a delay (phase difference) between the lag and flux waveforms, especially for non-QPO frequencies. Negative delays correspond to the flux waveform following the lag waveform, i.e. the lags vary first. Performing the same analysis on subsets of data ordered by their mean QPO frequency results in significant scatter between different data sets, especially in the phase difference between lag and flux, although almost all measured delays lie between 0 and -2~rad. Because the measured delays are difficult to understand, we investigated them using simulated light curves with constant lags, which is discussed in Subsection~\ref{subsection_comparison_sim}.

We confined the filter frequency in the FI method to between 0.6 and 2 times the QPO frequency, which is a slightly smaller range than for the filtered flux binning method introduced in Subsection~\ref{subsection_lag_var_timescales}. The FI method is slightly more sensitive to the Fourier leakage effects that show up in constant lag simulations, especially at higher frequencies, as is explained in Appendix \ref{app_rednoise}.

\subsection{Comparison with constant-lag simulated light curves}
\label{subsection_comparison_sim}
Because the process of dividing the light curves into short slices and binning them according to flux or QPO phase can introduce different kinds of Fourier leakage and systematic biases \citep{Uttley_2002,Alston_2013}, we tested our methods on simulated light curves as introduced in Appendix \ref{sim_appendix}. These simulated light curves have a constant lag, so they test the null hypothesis that the observed changing lags are systematic effects introduced by the methods we used. There are several meaningful similarities and discrepancies between the observations and the simulated data, which we highlight here.

First, it is important to note that none of the constant lag simulations can reproduce the linear relation between short-term time lags and (hard) flux, as shown in Fig. \ref{tlag_flux_alldata}, indicating that the observed relation cannot be due to the methods we used. We employed different methods of creating light curves consisting of a broadband noise signal multiplied\footnote{For completeness: we also tested the methods with simulations in which the QPO was added to the broadband noise or which did not contain a separate QPO signal. These simulations yielded similar results.} by a QPO signal and convolved with an impulse response function (see Appendix \ref{sim_appendix} for more on the simulations) and included the ubiquitously-observed rms-flux relation \citep{Uttley_2005,Heil_2012} by exponentiation of the original light curve. The result of applying the FI method to simulated data is shown in Fig. \ref{filterfreqfig_sim}, which shows for the simulated data the same measured parameters as Fig. \ref{filterfreqfig}, obtained from fitting single sine waves to the simulated lag and flux waveforms. Because the time lags are simulated in the Fourier domain, we expect them to stay constant in time. However, despite the lack of a direct relation between lag and flux, Fourier leakage effects can induce a waveform in the lags, especially at higher filter frequencies, albeit with a small amplitude (up to $\sim$0.1~ms) when compared to the data (up to $\sim$0.3~ms). The measured phase difference between the lag and flux waveform is not distributed uniformly, as would be the case if there were no Fourier leakage, but there is a clear preference for phase differences between -2 and -1~rad in the simulations, especially at higher filter frequencies. The negative phase differences signify that variations in flux follow variations in the lag, just like we observe in the data. By simulating light curves both with and without different rms-flux relations and for different lags, we found that the rms-flux relation is critical for understanding the effect.

Due to Fourier leakage, phase bins which correspond to a rising or falling part of the waveform (i.e. showing an overall trend), tend to have smaller (absolute) values for the measured lags, while peaks and troughs in the waveform do not suffer from leakage and have larger lags (see Appendix \ref{app_rednoise}). In simulated light curves without an rms-flux relation, the lag waveform due to leakage produces a harmonic signal with a low ($\lesssim$0.1~ms) amplitude. Implementing an rms-flux relation by exponentiating the light curve results in a lag waveform with a slightly higher amplitude, but more importantly, it introduces larger lags in the phase bins with a positive slope and lower lags in the phase bins with a negative slope. Due to this asymmetry, we measure a $\sim$ quarter-cycle delay of the flux compared to the lags when fitting single sine waves to both waveforms. To confirm that the rms-flux relation causes the delay, we also created light curves with an inverse rms-flux relation by subtracting the exponentiated light curve from a constant, increasing the amplitude of variability and finally setting all negative count rates equal to zero. The obtained light curves show a negative rms-flux relation and indeed we find that the lags follow the flux for these simulations, indicating that a larger rms increases the Fourier leakage and decreases the lags in subsequent phase bins.

The leakage effect described above could explain the lag vs. flux phase-difference behaviour that we see in the data, where the phase difference becomes more negative at non-QPO filter frequencies. The value of the delay is much closer to zero in the data than in the simulations, which is expected if the general lag-flux relation is (close to) instantaneous but Fourier leakage effects push it to negative values.
To better understand biases in the lag vs. flux phase difference, simulations which can reproduce an instantaneous linear change in lag with flux are required. The Fourier-based simulation methods we employ here are efficient, but they are limited by the fact that it is difficult to vary the lags over short time-scales. In order to introduce time-dependent lags, the light curves should be simulated in the time domain. This could be done with a more physics-based model, which we leave for future work. 

\section{Discussion}
We have shown that the short-term (4--20~Hz) 3--10~keV vs. 0.5--1~keV time lags, observed in the broadband noise of BHXRB MAXI~J1820+070, are linearly correlated with flux over a broad range of time-scales, with the strongest correlation seen between the lags and the harder, power-law dominated flux. Notably, the steepest lag vs. flux relation occurs on the QPO time-scale. The flux and lag waveforms on the QPO time-scale are similar, with only a small ($\sim0.2$~rad or $\sim0.03$~cycles, see Fig. \ref{filterfreqfig}) phase delay between them, which we attribute to bias caused by Fourier leakage effects (which probably also causes the delay between variations of lag and flux observed on other time-scales).

The linear lag-flux relation seems to be caused by the evolution of the lag vs. frequency dependence, with the cross-over frequency from hard to soft lags increasing with flux and the maximum soft lag decreasing at the same time, so that the net hard lag in the 4--20~Hz range increases with flux. Similar changes in the high-frequency lags are seen on much longer time-scales, as the source evolves through the hard state (e.g. as seen in Figure~\ref{ps_hs_lags_3sets} and shown by \citealt{Kara_2019}, \citealt{Wang_2021} and \citealt{De_Marco_2021}). \citet{Wang_2022} analysed the high-frequency soft lags in many black hole X-ray binaries that were observed by NICER and found that these lags evolve in a similar way in all systems during an outburst.

\citet{Wang_2021} model the lags in MAXI~J1820+070 using the \textsc{reltrans} reverberation model \citep{Mastroserio_2018,Ingram_2019,Mastroserio_2021} and find that the variations in soft lag at high frequencies can be explained by changes in coronal height. In their interpretation, larger heights produce greater light travel delays and correspondingly larger lags, with coronal height changing from $\sim30$ to $>300$~$R_{g}$ through the hard to soft state transition. For MAXI~J1820+070 \citep{Wang_2021} and BHXRB GX~339-4 \citep{Wang_2020} the required large coronal heights appear to be in tension with results from spectral modelling of relativistic reflection, which imply more compact coronae. Note that in a recent work \citet{Lucchini_2023} reconcile both reflection and timing results by including two lampposts to simulate a more vertically extended corona. In the context of these more vertically-extended models, it is important to realise that the first polarisation results from IXPE for Cyg X-1 in the hard state and for Swift J1727.8-1613 in the HIMS favour a more horizontally extended corona \citep{Krawczynski_2022,ingram2023tracking}.

The observations we study here correspond to a relatively small range of lag variation during the bright hard state (compared to state changes), when the corona is likely more compact, at most a few tens of $R_{g}$ in height \citep{Wang_2021}. Coronal heights may be significantly smaller than derived from light travel times if disc mass-accretion propagation and seed photon effects are taken into account \citep{uttley2023large}. In any case coronal height changes may be required to explain the pattern of short-term lag variability seen in MAXI~J1820+070, with the coronal height \textit{decreasing} with increasing X-ray flux. The X-ray flux variations on non-QPO time-scales are thought to be linked to mass accretion fluctuations, which would further suggest that decreases in coronal height are linked to increases in mass accretion rate. 

Following the idea that the observed changes in timing properties are due to a variations in coronal geometry, we propose to distinguish two types of geometric change: observer-dependent and intrinsic. If the geometric change is observer-dependent, this means that e.g. the angular size or beaming of the coronal emission towards the observer is different, but the actual shape and emission of the corona stays constant. In the case of a precessing corona, its orientation shifts over time, causing variations in the emission towards the observer due to a combination of solid-angle changes, relativistic beaming effects and angular-dependence of coronal emission \citep{VeledinaQPO_2013}. These are observer-dependent variations. Because we do not expect changes in e.g. orientation angles over a time-scale of weeks, this type of geometric variability is only feasible over short time-scales, such as due to a precessing hot flow or jet in the Lense-Thirring framework. 

We define intrinsic coronal changes as variations in geometry that do influence the shape and/or emission of the corona, independent of the observer's viewing angle. The shape or size of the corona actually change for this type of geometric change, instead of only the coronal orientation, as is the case in observer-dependent changes. To give an example, an increase in coronal height can take place over a broad range of time-scales, e.g. as the accretion rate changes during an outburst, but probably also on a time-scale of seconds. The increased coronal height will affect the seed photon flux towards the corona, changing spectral-timing properties.
Long-term lag differences, such as presented in \citet{Kara_2019}, \citet{De_Marco_2021} and \citet{Wang_2022} will thus be due to intrinsic rather than observer-dependent changes in coronal geometry. Lense-Thirring precession also includes an intrinsic geometric change, in that it also causes the hot flow to \emph{nutate}, i.e. there is a (quasi-periodic) change in the polar angle of the precessing hot flow. From the point of view of the disc, the height of one side of the precessing flow is varying at the QPO frequency, which affects the seed photon flux coming from the disc and associated spectral and timing properties.

As discussed in Section~\ref{sec:intro}, there is substantial evidence that QPO timing properties depend on binary orbit inclination \citep{Motta_2015,Heil_2015,van_den_Eijnden_2016,de_Ruiter_2019}, which in turn suggests that the QPO corresponds to some kind of changing inner-region geometry. It is possible that the QPO (in the flux) is itself produced by the same variations in coronal geometry (e.g. optical depth and corresponding height changes due to accretion rate fluctuations) that produce the lag variations and which are somehow enhanced at the QPO frequency. For example, if an increase in X-ray flux corresponds to a shrinking corona, then the QPO may correspond to a particular frequency at which the coronal height oscillates. This picture is inconsistent with the precessing hot flow model for the QPO however, since in that case the quasi-periodic flux variation at a particular frequency is linked to the precession of a hot flow with otherwise constant shape. 

In the precession model, lag changes might be produced due to e.g. light-travel delay variations as the hot flow preferentially illuminates the near and far side of the disc (as seen by the observer) at different precession phases (see e.g. \citet{Ingram_2012,You_2018} and \citet{You_2020} for examples of varying coronal illumination of the disk, but note that none of these papers discuss the lags that are the focus of the current paper). Including nutation in the model could mimic coronal height changes at the QPO frequency. Coronal precession at the QPO frequency and coronal height changes linked to accretion rate fluctuations on a broad range of time-scales could then happen simultaneously and potentially explain our obtained results, as they would both correspond to changes in solid angle as seen from the disc and from an observer. 

\begin{figure}
    \centering
    \includegraphics[width =\linewidth]{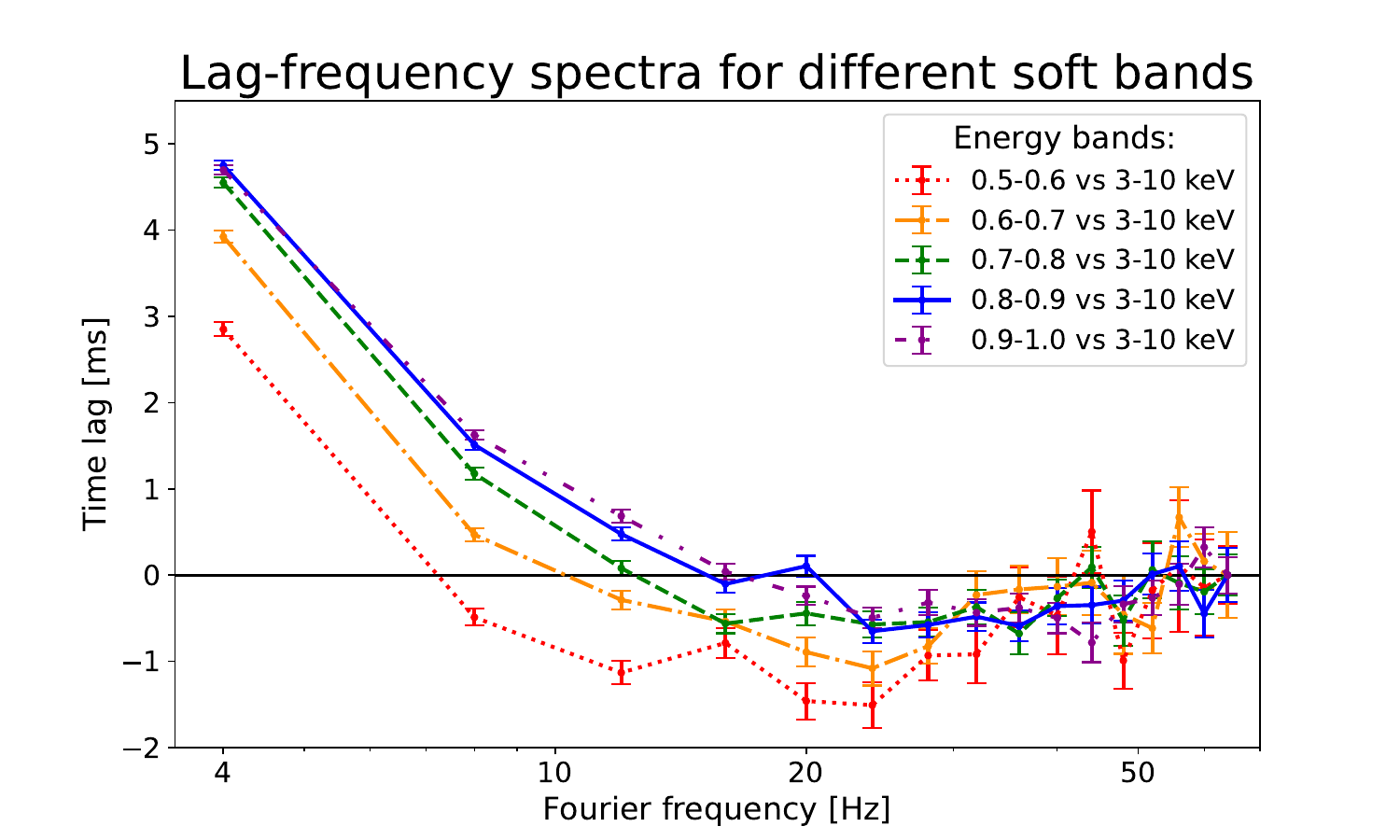}
    \caption{The time lag vs frequency spectra for all data combined, using five different soft bands between 0.5 and 1~keV. The lag depends strongly on the choice of soft band energies and the difference is reminiscent of Fig. \ref{tlag_flux_alldata}, where the soft band is 0.5--1~keV but the data are split into five different flux bins.}
    \label{lagfreq_5soft}
\end{figure}

To understand the changing short-term lags at both QPO and broadband-noise-dominated time-scales, we also looked the energy-dependence of the time lags, which is demonstrated in Fig. \ref{lagfreq_5soft}. In the figure, we show the time lag vs frequency spectra for narrow soft bands between 0.5 and 1~keV, which can be compared to Fig. \ref{tlag_flux_alldata}. The lag behaviour for slightly different soft bands is similar to what we see when binning on hard flux and the range of time lags covered is also comparable, so spectral changes might explain at least part of the lag vs flux relation. Reverberation lags are observed when comparing disc and power-law dominated bands, while the lags between different power-law energies do not become negative and are generally positive or consistent with zero lags at high Fourier frequencies. Spectroscopic fits show that the 0.5--1~keV energy band we use here consists of  photons from the disc (between 50 and 70\% for all observations) and from the corona (50 to 30\%)\footnote{These values are the result of fitting spectra in the 0.3-10 keV range with two models. The first spectral model, \texttt{tbabs*simpl*diskbb}, corresponds to the situation that the seed photons of the Comptonization component originate from the entire accretion disc, yielding a lower limit for the 0.5--1~keV disc contribution. The second model is \texttt{tbabs*(simpl*bb+diskbb)}, which corresponds to seed photons only arising from the inner parts of the disc and provides an upper limit for the soft band disc contribution. For all studied observations, the obtained values for the fraction of photons from the disc lie between 50 and 70\%.}. A change in power-law normalisation could impact the weighting of both components and their associated lags (i.e disc vs power-law and medium power-law vs hard power-law). For example, a larger normalisation of the power-law would lead to measuring harder lags as the power-law vs power-law lags play a larger role. We observe a similar pattern in the data, which in this view does not require the intrinsic disc vs power-law lags to change as much. It remains to be explained why the relative strength of the spectral components in the 0.5--1~keV energy range changes, which could still be due to a change in geometry.

From QPO resolved spectroscopy, we know that the QPO is associated with a strong modulation of the power-law normalisation \citep{Ingram_2016_IronK}. A recent paper by \citet{Gao_2023} reached a similar conclusion for the hard state QPOs in MAXI~J1820+070 with {\it Insight-HXMT} data. The large power-law flux modulation at the QPO frequency would explain why the largest change in short-term time lags happens at the QPO frequency. The normalisation is expected to change on other time-scales as well, providing an explanation for the measured lag vs flux relation on broadband noise frequencies. The idea that the main spectral property of the QPO is a varying power-law normalisation, with a small change in photon index $\Gamma$, is consistent with a geometric origin \citep{You_2020}. The coronal emission would not change (much) intrinsically, but a change in solid angle and relativistic boosting effects result in a modulation of the power-law flux. The same would go for e.g. coronal height changes linked to accretion rate fluctuations. Time- and energy dependent simulations can shed light on the effects of changing the strength of different spectral components on the measured time lags. 

The interpretation of the lag vs flux relation in this section depends strongly on the model used to explain the soft disc vs power-law lags routinely observed in BHXRBs, which we assume to be reverberation of power-law photons on the disc. Other models have been proposed (see e.g. \citealt{Mushtukov_2018} and \citealt{Kawamura_2023}) and those could lead to different conclusions. Because the measured short-term time lags arise due to several different mechanisms (e.g. reverberation, power-law pivoting), changes in the hard lags within the power-law component at those frequencies would cause similar observational results. However, these hard lags are probably also strongly affected by the geometry so they do not change our overall conclusion. Again, more detailed physical simulations could shed more light on the effects of a change in lags in different spectral components.

MAXI~J1820+070 is a remarkably bright source, which spent a large amount of time in the hard state so that a detailed study of the kind presented here could be carried out using NICER's exceptional data. Similar studies of other BHXRBs will be more challenging but could shed important light on whether the observed lag variability is common or specific to MAXI~J1820+070 or sources like it. E.g. the lag variability may also depend on system inclination, which might be expected given the enhanced lag variability at the frequency of the (inclination-dependent) QPO signal.  

In our analysis, we focused on changes in the short-term time lags on the QPO time-scale and broadband-noise-dominated time-scales. If the fluctuating short-term lags can be attributed to geometric variations, our results sketch a picture of a dynamic inner X-ray emitting region. Assuming this, the strongest change in geometry is observer-dependent and takes place on the QPO time-scale, while slightly smaller intrinsic variations happen on both longer and shorter time-scales and are linked to accretion variability. We can connect these findings to recent GRMHD simulations of accreting black holes by e.g. \citet{Liska_2018}, \citet{Liska_2022} and \citet{Musoke_2022}, which show that the component that could serve as the corona is far from being a static region. The simulations include a strong and dynamic magnetic field, which could be key to understanding coronal geometry variations on short time-scales.

\section{Conclusions}
Our main findings can be summarised as follows:
\begin{enumerate}
    \item The short-term (4--20~Hz) time lags between 0.5--1~keV and 3--10~keV light curves show a positive linear relation with the hard flux.
    \item The relation between short-term time lags and hard flux exists for variability on a range of time-scales and is strongest at the QPO frequency.
    \item In the framework of X-ray reverberation, we interpret the lag-flux coupling as being due to geometric changes on time-scales of seconds to tens of seconds, with the strongest variations taking place on the QPO time-scale.
\end{enumerate}
We have introduced our new filter-interpolate method, which is designed to perform QPO phase-resolved spectral-timing. We tested our method both by comparing its results with those from a more established method (Appendix \ref{appendix_waveform}) and also by applying it to simulated `null hypothesis' light curves with constant lags (Appendix \ref{sim_appendix}). Although these simulated light curves can replicate some important properties of both the QPO and the broadband noise in real data, they cannot reproduce the short-term time lag - hard flux relation. In order to simulate the lag - flux relation, more physical time- and energy dependent light curve simulations are required, which can also be used to study the effects of changing the strength of different spectral components. Our results support a geometric origin of low frequency QPOs, but also suggest that the corona is a dynamic structure over a broad range of time-scales, whose variable nature should be taken into account by any explanatory model.

Because the spectral-timing properties in BHXRBs vary on short time-scales, we should be somewhat cautious when interpreting time-averaged spectral-timing results. Large future X-ray observatories, such as \emph{Athena} \citep{ATHENA_2013}, \emph{eXTP} \citep{eXTP_2016} and \emph{STROBE-X} \citep{STROBE-X_2019}, could address this issue by resolving spectral-timing variations down to short time-scales for a much larger sample of sources, providing important information on the dynamics of the corona.

\section*{Data Availability}
 This research has made use of data obtained through the High Energy Astrophysics Science Archive Research Center Online Service, provided by the NASA/Goddard Space Flight Center. The data underlying this article are available in HEASARC, at \url{https://heasarc.gsfc.nasa.gov/docs/archive.html}. A basic reproduction package for the results and figures in this paper is available on Zenodo via \url{https://doi.org/10.5281/zenodo.10391399}.

\section*{Acknowledgements}
The authors thank the anonymous referee for their helpful comments. 
JvdE acknowledges a Warwick Astrophysics prize post-doctoral fellowship made possible thanks to a generous philanthropic donation and was supported by a Lee Hysan Junior Research Fellowship awarded by St Hilda's College, Oxford during part of this work. AZ is supported by NASA under award number 80GSFC21M0002.\\
This research makes use of the SciServer science platform (\url{www.sciserver.org}). SciServer is a collaborative research environment for large-scale data-driven science. It is being developed at, and administered by, the Institute for Data Intensive Engineering and Science at Johns Hopkins University. SciServer is funded by the National Science Foundation through the Data Infrastructure Building Blocks (DIBBs) program and others, as well as by the Alfred P. Sloan Foundation and the Gordon and Betty Moore Foundation. \\
This work made use of Astropy:\footnote{http://www.astropy.org} a community-developed core Python package and an ecosystem of tools and resources for astronomy \citep{astropy:2013, astropy:2018, astropy:2022}. Also, we made extensive use of NumPy \citep{harris2020array} and SciPy \citep{2020SciPy-NMeth}.
 


\bibliographystyle{mnras}
\bibliography{short} 


\newpage
\appendix

\section{Simulations}
\label{sim_appendix}
\label{section_simulations_waveform}
We tested our methods using "null hypothesis" simulated light curves with a constant lag. The broadband noise is simulated following the method of \citet{TK95}. We used the multi-Lorentzian fit of the hard band power spectrum of a representative observation, ObsID 144, taken on May 4, 2018, as input for the simulation power spectrum. We removed the narrow Lorentzian components which fitted the QPO fundamental and harmonic. The observed shape of the broadband noise power spectrum in the data is approximated well in this way.

\begin{figure}
    \centering
    \includegraphics[width=\columnwidth]{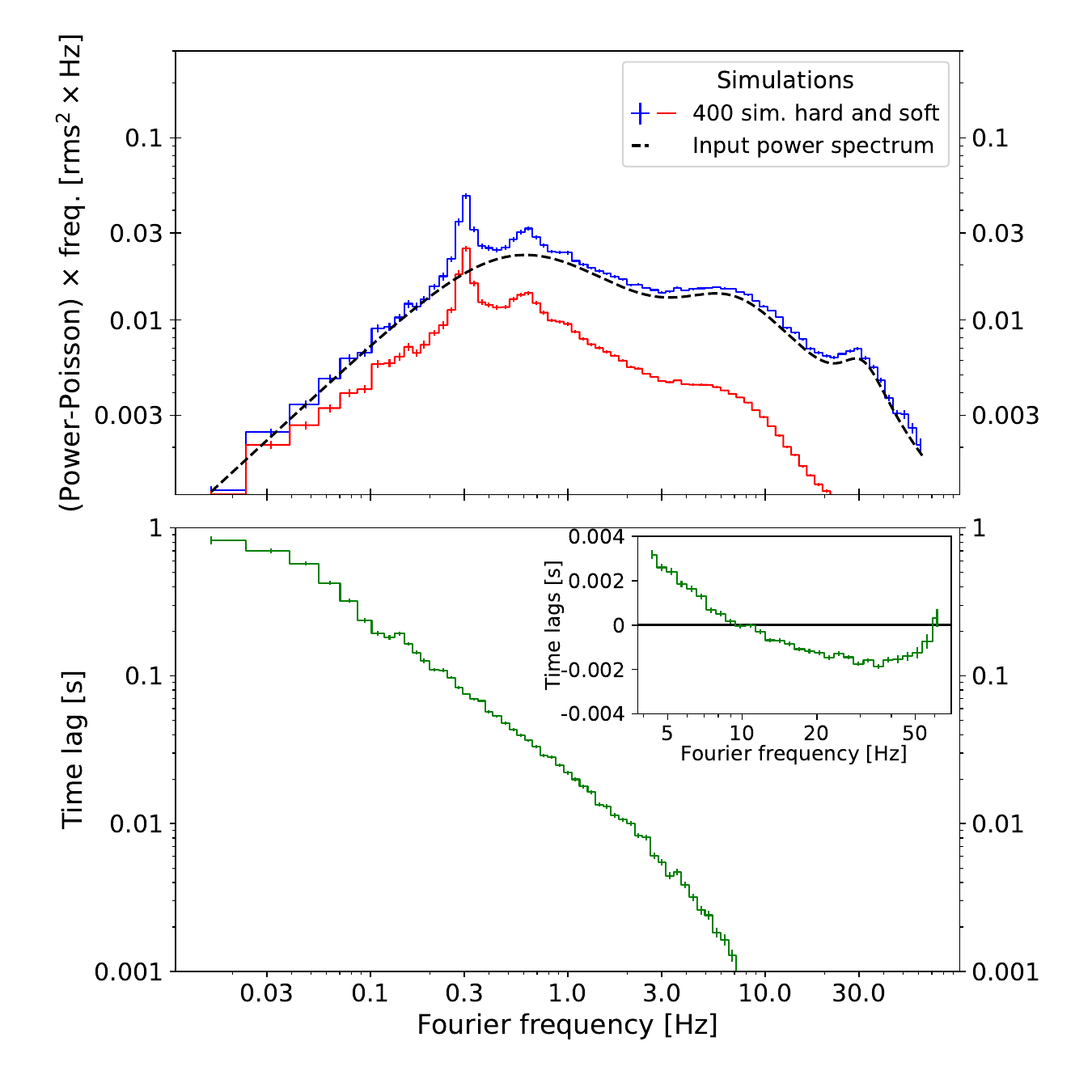}
    \caption{The upper panel shows two power spectra. The upper blue spectrum is created by taking the average over 400 simulated segments of 64 seconds. It is very similar to the input power-spectral shape, which is based on the three Lorentzians fitting the broadband noise of ObsID 144, shown as the black dashed line, except for the peaks that resemble QPOs, which are due to multiplication with a simulated QPO signal. At high frequencies, there is a small positive offset from the input power spectrum, which is the result of exponentiating the light curve to include the rms-flux relation. The contribution of Poisson noise to the power spectrum has been subtracted. The lower red power spectrum is obtained from convolving the `hard' light curve with the impulse response shown in Fig. \ref{IRF_BBN}. The lower panel shows the resultant lag-frequency spectrum. The simulations in this figure can be compared to the data in Fig. \ref{ps_hs_lags_3sets}, which look very similar. The light curve corresponding to the `hard' power spectrum is visible in the lower panel of Fig. \ref{ampmodbbn}. }
    \label{pslags_sim}
\end{figure}

The original simulated light curve has mean zero and we obtain the final light curve by exponentiating it, $x_{\rm e}(t) = \rm{e}^{\it{x}_{\rm{n}} \it{(t)} }$, where $x_{\rm e}(t)$ is the exponentiated light curve and $x_{\rm n}(t)$ the original (normally distributed) light curve. Exponentiation prevents light curve values from dropping below zero (which is unphysical) and accounts for the rms-flux relation, which was observed to be ubiquitously present by \citet{Heil_2012} and which is naturally recreated by exponentiation \citep{Uttley_2005}. The exponentiation does have a small effect on the width of the power spectrum, which is just visible in Fig. \ref{pslags_sim}. The broadening effect can be significant when the model PSD contains sharp features, but for the broad power distributions we used, the effect is negligible. 

The QPOs are simulated using a different approach. The most important aspect is that the QPO signal is interpreted as the sum of two phase-linked sine waves with a phase offset that follows a random walk, which was also done by \citet{Krawczynski_2016} and \citet{ingram2020review} and is described by
\begin{equation}
    x_{\textsc{qpo}}(t) = 1 + A \cos\left(2\pi t \nu_{\rm c} - \phi_{\rm rw}(t)\right) + \\ \\
    B \cos\left(2\left[2\pi t \nu_{\rm c} - \phi_{\rm rw}(t)-\Psi\right]\right).
    \label{qpo_sim_eq}
\end{equation}
In this equation, $x_\textsc{qpo}(t)$ is the QPO signal, $A$ and $B$ are the amplitudes of the fundamental and the harmonic respectively, $\nu_{\rm c}$ is the fundamental centroid frequency, $\Psi$ is the constant phase difference between the fundamental and harmonic phase which defines the waveform shape (see Equation~\ref{equation_psi}), and $\phi_{\rm rw}$ is the random walk phase offset of the fundamental. 
The random walk is created by $\phi(t_i)=\phi(t_{i-1})+\epsilon_i$, where $\epsilon_i\sim\mathcal{N}(0,\sigma^2)$. $\mathcal{N}$ is a normal distribution with mean 0 and standard deviation $\sigma$ equal to the step size of the random walk, which determines how fast the QPO signal goes out of phase with a purely periodic signal. In other words, it determines the width of the QPO peak in the power spectrum, the Q-value. In the figure, the step size is 0.04, while the time resolution is 1/128 s and the QPO frequency 0.3 Hz, corresponding to $\text{Q}\sim6$. The random walk starts between 0 and $2\pi$.

\begin{figure}
    \includegraphics[width=\columnwidth]{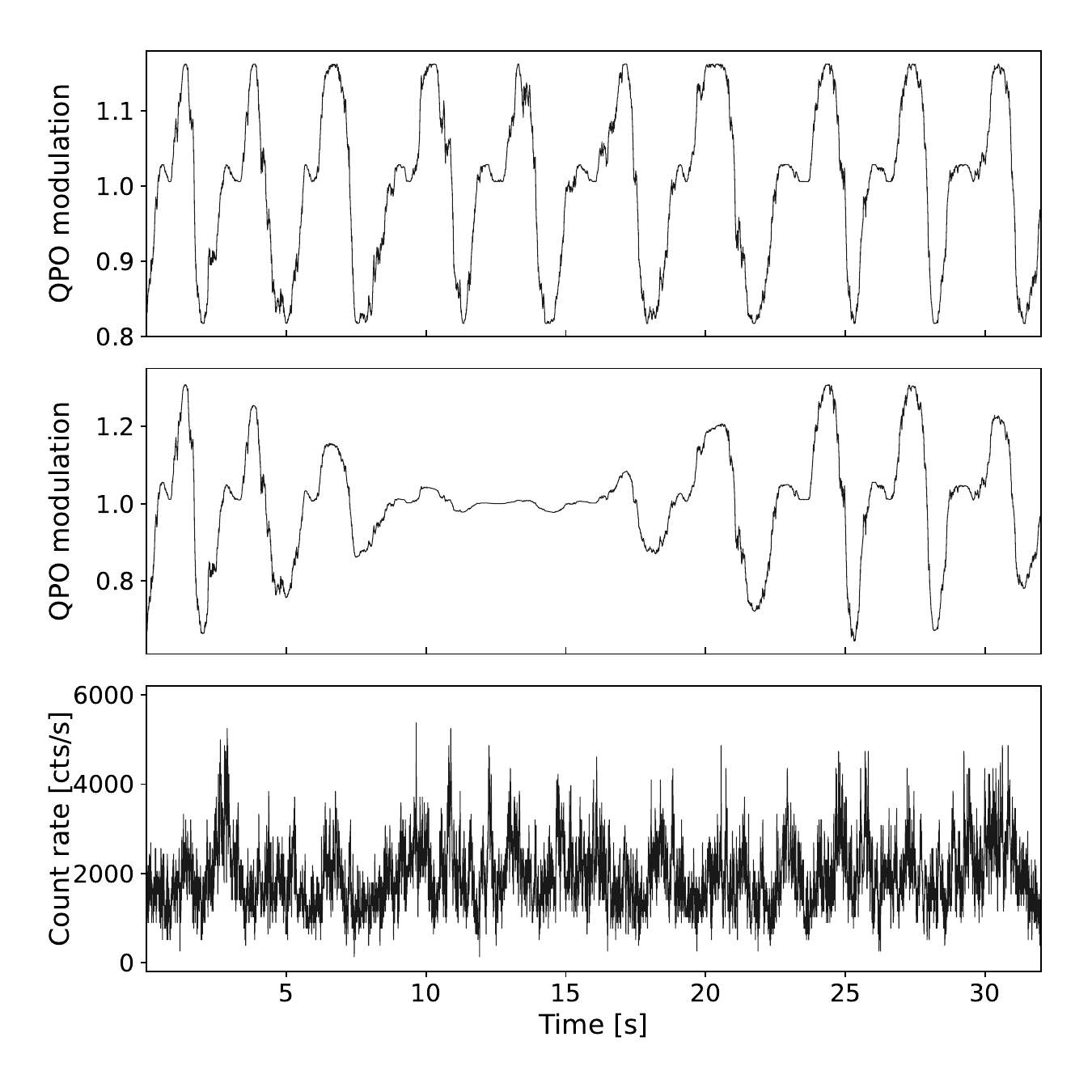}
    \caption{Three simulated light curves with a QPO signal. The upper panel shows the QPO component without amplitude modulation, the middle panel contains a simulation with amplitude modulation. The random walk step size used here is 0.04, corresponding to $\rm{Q}\approx6$ for $\nu_{\rm{QPO}}=0.3$ Hz and 1/128 s time bins. $\Psi=0.3$ rad/$\pi$ in both cases, which is similar to the value found in the data, and the amplitude of the fundamental is 1.5 times the amplitude of the harmonic. The product of the constant amplitude QPO component and a broadband noise component is shown in the lower panel, where the light curve has been scaled to realistic values and also includes Poisson noise.}
    \label{ampmodbbn}
\end{figure}

The result of equation \ref{qpo_sim_eq} is a double sine waveform with a fluctuating phase (or equivalently, frequency). The phase-linked harmonic signal at twice the centroid frequency enables us to simulate QPO waveforms that are more complicated than a single sine wave, which have been observed in many black hole X-ray binaries \citep{de_Ruiter_2019}. An example of such a waveform is shown in the upper panel of Fig. \ref{ampmodbbn}, with phase difference $\Psi=0.3\pi$ rad.

Both amplitudes $A$ and $B$ can be made dependent on time, $A(t)$ and $B(t)$, to account for the fact that they might vary in observed QPOs as well. \citet{van_den_Eijnden_2016} found that in GRS 1915+105, coherent intervals exist for the QPO signal that seem to be connected to the Q-value. In our simulation, $A(t)$ and $B(t)$ can be described by another quasi-periodic sine wave with a mean period equal to the length of coherent intervals. An example of what this looks like can be seen in the middle panel of Fig. \ref{ampmodbbn}. In the rest of this paper, we only show results with a constant QPO amplitude (before being modulated by the broadband noise, see below), mainly because amplitude modulation is an extra complication that does not change our conclusions for the types of tests we conducted.

We obtain the final combined light curve by multiplying (as opposed to adding) the generated broadband noise by the QPO signal, which is more consistent with the QPO being due to geometric change and with the observed coupling of QPO and broadband noise signals (e.g. \citealt{Heil_2010, Maccarone_2011, Arur_2019}). The mean of the QPO signal is in this case normalised to 1, with the sum of the fundamental and harmonic amplitudes $<1$, in order to prevent negative values. The Fourier transform of two multiplied signals is the convolution of their respective Fourier transforms. The final product of simulating light curves by multiplying the QPO and broadband noise components, scaling to the required mean count rate and including Poisson noise is shown in the lower panel of Fig. \ref{ampmodbbn}. 

The methods for QPO phase-resolved spectral-timing require two correlated light curves from different energy bands with frequency-dependent time lags. We use impulse response functions (henceforth ‘impulse response’) to introduce these lags between two simulated light curves (e.g. \citealt{Uttley_2014}). In our simulations, we first simulate the broadband noise and QPO signal of the `hard band', because the convolution with the impulse response will lead to a reduction of power, especially in higher frequencies. Fig. \ref{ps_hs_lags_3sets} clearly shows that the soft band fractional rms power is lower than the hard power spectrum at almost all frequencies in the hard state of MAXI J1820+070. This is also visible in the upper panel of Fig. \ref{pslags_sim}, where we show the input power spectrum and the simulated `hard' and `soft' bands, which can be compared to the upper panel of Fig. \ref{ps_hs_lags_3sets}. 
\begin{figure}
    \centering
    \includegraphics[width=\columnwidth]{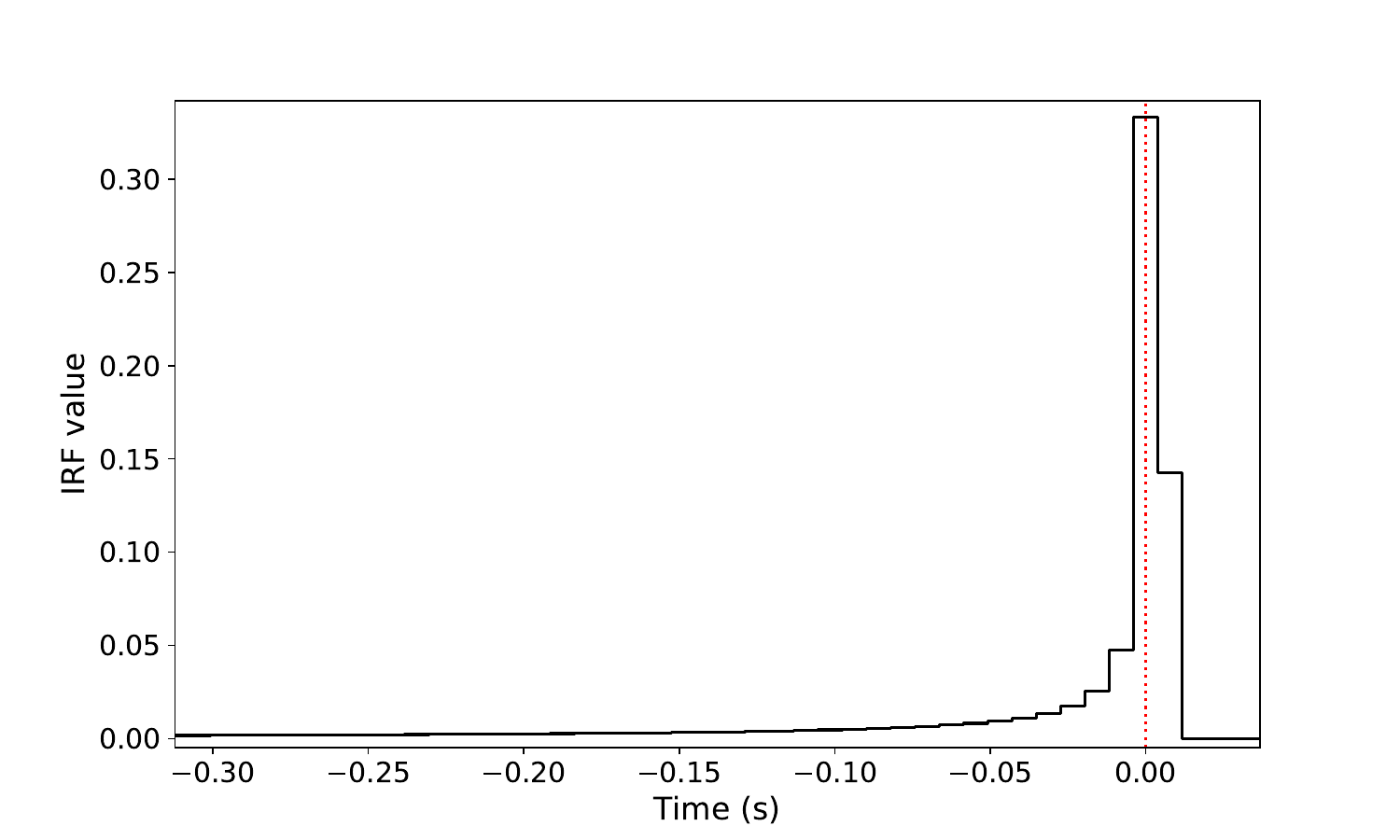}
    \caption{The impulse response used for simulations of broadband noise. If this impulse response is applied to the simulated hard band, the hard band will lag behind the soft band at low frequencies (due to the broad structure at negative times), while the sign of the lag is reversed for high frequencies, due to the narrow function at positive times. The dotted red line indicates zero delay. The impulse response has already been binned to the light curve time resolution.}
    \label{IRF_BBN}
\end{figure}

To simulate the `soft band', the impulse response in Fig. \ref{IRF_BBN} is convolved with the simulated `hard' light curve, which consists of both the broadband noise and the QPO. We use an empirical form of the impulse response given by
\begin{equation}
  f(t) =
    \begin{cases}
    \rm{c_0}|t|^{-\alpha} &\text{for } -t_s<t<0 \\
    \rm{c_1}              &\text{for } t=0 \\
    \rm{c_2}              &\text{for } t=\delta t \\
    0                     &\text{for } t<-t_s \lor t>\delta t ,
    \end{cases}  
\end{equation}
where $\rm{c_0}$ is a normalising constant, power-law index $\alpha=0.9$, $t_s=10$ s is the longest time-scale for which we define lags, $\rm{c_1}$ and $\rm{c_2}$ are constants determining the amplitude of all lags and soft lags, respectively, and $\delta t$ is the time resolution of the light curve. 
The power-law results in hard lags at low frequencies and the narrow positive component causes the soft lags at high frequencies. The resultant lag-frequency spectrum is very similar to those obtained from the data, as is visible in the lower panel of Fig. \ref{pslags_sim}, which can be compared to the lower panel of Fig. \ref{ps_hs_lags_3sets}.

\section{Testing the filter-interpolate method with QPO waveform reconstruction}
\label{appendix_waveform}
To test the filter-interpolate (FI) method presented in Section~\ref{subsection_qpophase_sawtoothmethod}, we used it to reconstruct the QPO waveform (modelled using Equation~\ref{equation_qpo_main}) and compare the obtained phase difference between fundamental and harmonic $\Psi$ to the value obtained by another method, designed by \citet{Ingram_2015} and applied to many systems by \citet{de_Ruiter_2019}. We will refer to the latter approach as the Fourier phase offset (FPO) method from now on. It works by splitting the light curve into many small segments and then estimating the waveform phase difference $\Psi$ from the statistical distribution of observed Fourier phase differences between the fundamental and harmonic frequencies. For a more extensive explanation of the FPO method and its use, we refer to \citet{Ingram_2015}, \citet{Ingram_2017} and \citet{de_Ruiter_2019}. MAXI J1820+070, has not been studied using the FPO method before. For both methods, errors were determined by bootstrapping. More recently, \citet{Nathan_2022} introduced a more sophisticated method to calculate the phase difference between the fundamental and the harmonic using the bispectrum. We also applied their method and obtained very similar results.

\begin{figure}
    \centering
    \includegraphics[width=\columnwidth]{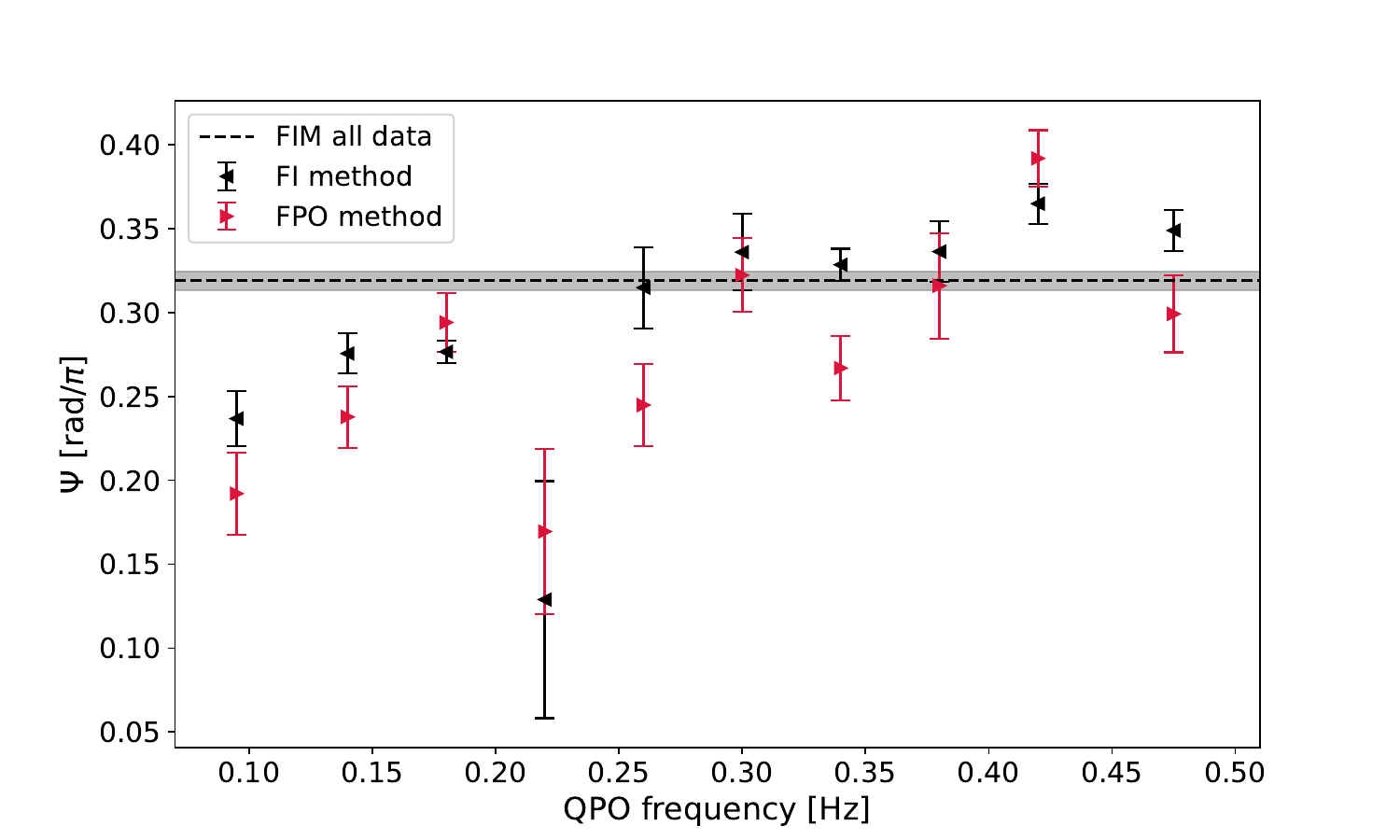}
    \caption{The phase difference between the fundamental and the harmonic for all data sets as defined in Table \ref{ObsID_datasets}. The x-axis value is the mean QPO frequency of each data set. The large error bars with a QPO frequency between 0.2 and 0.24 Hz arise due to the small amount of lower quality data in this range. It is clear both methods return comparable values for the phase difference between the fundamental and the harmonic. The mean phase difference $\Psi$ when combining all data as measured with the interpolation method is shown as the dashed line.}
    \label{phdiff_2methods}
\end{figure}

We applied the FI and FPO methods to the ten QPO frequency-selected data sets listed in Table~\ref{ObsID_datasets}, to obtain the phase difference for the fundamental vs. harmonic $\Psi$. The results are plotted in Fig. \ref{phdiff_2methods}. There is a significant ($3.7\sigma$ for the FI method and $>5\sigma$ for the FPO method) rise in the phase difference $\Psi$ with QPO frequency. The measured value of $\Psi\sim0.3$ rad/$\pi$ is reasonably consistent with the results from \citet{de_Ruiter_2019} for high inclination sources like MAXI J1820+070, even though the general trend for these sources is that the phase difference $\Psi$ decreases with QPO frequency. In their results, however, the scatter is large and the range of QPO frequencies studied in our work, 0.07 to 0.51 Hz, is modest compared to the total range covered by \citet{de_Ruiter_2019}. The phase difference values measured with both methods are consistent given the statistical errors\footnote{Using Pearson's $\chi^2$-test to compare the results from both methods for all data sets, we obtain a probability that the differences are due to statistical fluctuations of 0.007 ($2.7\sigma$). The low probability arises due to the relatively small error bars obtained by bootstrapping for some of the data sets.}, which gives us confidence that the FI method is correctly resolving the QPO signal, with no bias in the resulting waveform shape.

\section{Red noise leakage}
\label{app_rednoise}
The frequency range for which filtered light curves were created in section \ref{subsection_lag_var_timescales} is limited by red noise leakage, which creates a bias \citep{Alston_2013}. In order to mitigate the effect of the bias, we only made filtered light curves between 0.4 and 2.5 times the QPO frequency of each observation, corresponding to a maximum central filter frequency of $\sim$1.25 Hz. For the interpolation method, the filter range was even smaller, ranging from 0.6 to 2 $\nu_\textsc{qpo}$. In section \ref{subsection_comparison_sim}, the consequences of the Fourier leakage effects described below for the lag waveform are reported in more detail.

The bias can be understood by thinking of the average shape of the light curve slices in each bin. When binning short light curve slices by their filtered flux, such as in section \ref{subsection_lag_var_timescales},  the slices with the lowest and highest filtered flux will on average contain a peak or trough in the flux and will not contain any strong overall flux-trend. The light curve slices in other bins will on average contain a slope that is due to variability at frequencies below the sampled Fourier frequency range, which induces leakage of these lower frequencies and decreases the measured (absolute) time lags, since the trend will appear the same in both bands. 

The rms-flux relation affects the leakage in such a way that the highest filtered flux bin has larger absolute lags than the lowest flux bin. The end result of the different light curve shapes in each flux bin is that there seems to be a linear relation between the short-term time lags and the hard flux, at least when the period of the filter is not at least a few times larger than the slice length (0.25~s). At the same time, light curves filtered at high frequencies are not well-correlated with the total flux, leading to only a small range in hard flux. The combination of a modest range in lags and a small hard flux range introduces steep slopes. Simulated light curves show very similar results when using filtered light curves at high Fourier frequencies (>2 Hz for 0.25 s slices), while this is not the case for lower filter frequencies, indicating that the slopes arise due to systematics. We therefore do not show results for these higher frequency filters in this paper.

\section{Observation list}
\label{appendix_obslist}
\begin{table*}
    \centering
    \scriptsize
    \begin{tabular}{c|c|c|c|c}
        \hline
       ObsID &$\nu_\textsc{qpo}$ [Hz] & Sel. exp. time [s]& Flux 0.5-1 keV [cts/s] &  Flux 3-10 keV [cts/s]\\
       \hline\hline
        126 & 0.077 & 1280 & 7143 & 1350 \\
        127 & 0.082& 640 & 7011 & 1306 \\
        129 & 0.094 & 448 & 7212 & 1338 \\
        130 & 0.12 & 3712 & 6863 & 1303 \\
        131 & 0.12 & 1920 & 6141 & 1159 \\
        132 & 0.14 & 1920 & 7029 & 1310 \\
        133 & 0.14 & 1792 & 6972 & 1292 \\
        134 & 0.14 & 4928 & 7027 & 1307 \\
        135 & 0.17 & 3200 & 7118 & 1311 \\
        136 & 0.17 & 1344 & 7169 & 1304 \\
        137 & 0.17 & 5760 & 6963 & 1264 \\
        138 & 0.18 & 3264 & 6962 & 1260 \\
        139 & 0.17 & 512 & 6896 & 1244 \\
        140 & 0.25 & 2176 & 7187 & 1250 \\
        141 & 0.27 & 832 & 7132 & 1211 \\
        142 & 0.28 & 4800 & 7029 & 1184 \\
        143 & 0.29 & 3648 & 7030 & 1167 \\
        144 & 0.31 & 4608 & 7083 & 1170 \\
        145 & 0.34 & 5888 & 7149 & 1159 \\
        146 & 0.35 & 4928 & 7064 & 1142 \\
        147 & 0.39 & 4544 & 7272 & 1146 \\
        148 & 0.41 & 3776 & 7374 & 1142 \\
        149 & 0.45 & 2048 & 7517 & 1149 \\
        150 & 0.46 & 1600 & 7581 & 1130 \\
        151 & 0.42 & 704 & 7250 & 1119 \\
        152 & 0.46 & 1984 & 7356 & 1101 \\
        153 & 0.46 & 768 & 7465 & 1104 \\
        154 & 0.50 & 448 & 7520 & 1084 \\
        155 & 0.51 & 576 & 6899 & 1008 \\
        156 & 0.49 & 4800 & 6758 & 980 \\
        157 & 0.41 & 4608 & 6158 & 931 \\
        158 & 0.35 & 2176 & 5701 & 891 \\
        159 & 0.34 & 1664 & 5477 & 859 \\
        160 & 0.34 & 2048 & 5224 & 824 \\
        161 & 0.34 & 3264 & 5260 & 827 \\
        162 & 0.34 & 2240 & 5147 & 807 \\
        163 & 0.42 & 2816 & 5466 & 821 \\
        164 & 0.41 & 1856 & 5316 & 803 \\
        165 & 0.38 & 2624 & 5034 & 774 \\
        166 & 0.37 & 1984 & 4907 & 756 \\
        167 & 0.35 & 1856 & 4752 & 732 \\
        168 & 0.35 & 960 & 4699 & 729 \\
        169 & 0.31 & 1664 & 4242 & 669 \\
        170 & 0.34 & 3712 & 4302 & 668 \\
        171 & 0.32 & 2176 & 4011 & 630 \\
        172 & 0.35 & 2112 & 3996 & 617 \\
        173 & 0.28 & 3008 & 3642 & 576 \\
        174 & 0.28 & 2560 & 3427 & 546 \\
        175 & 0.31 & 2112 & 3260 & 516 \\
        176 & 0.27 & 5696 & 3097 & 490 \\
        177 & 0.24 & 704 & 2984 & 469 \\
        178 & 0.27 & 2432 & 2732 & 436 \\
        179 & 0.31 & 832 & 2662 & 420 \\
        180 & 0.25 & 2112 & 2536 & 403 \\
        181 & 0.24 & 384 & 2251 & 373 \\
        182 & 0.22 & 1856 & 2252 & 365 \\
        183 & 0.24 & 640 & 2262 & 367 \\
        184 & 0.24 & 640 & 2246 & 363 \\
        185 & 0.22 & 1728 & 2224 & 364 \\
        186 & 0.27 & 1664 & 2669 & 419 \\
        187 & 0.29 & 1280 & 2891 & 450 \\
        188 & 0.36 & 1536 & 3754 & 567 \\
        189 & 0.42 & 12800 & 4226 & 615 \\
        \hline
    \end{tabular}
    \caption{The ObsIDs used in this paper and the observation time obtained by excluding GTIs of less than 64 seconds. Also, the average QPO fundamental frequency in each observation and the mean count rates for the soft 0.5-1 and hard 3-10 keV energy bands are shown. The total observation time is 158,592 s.}
    \label{ObsID_list}
\end{table*}

\label{appendix}



\bsp	
\label{lastpage}
\end{document}